\documentclass[12pt]{iopart}
\usepackage{graphicx}
\begin{document}

\title[Transport Channels in a Double Junction]
{Transport Channels in a Double Junction -
Coherent coupling changes the picture}

\author{U Schr\"oter and E Scheer}
\address{University of Konstanz, FB Physik,
Universit\"atsstra\ss e 10, 78464 Konstanz, Germany}

\ead{Ursula Schroeter@uni-konstanz.de}

\begin{abstract}
Transport through a point contact is accurately modelled by assigning
to the junction an ensemble of independent transport channels with possibly
different transmissions. We here argue that for a series of two contacts,
coherently coupled across an island,
the transport channels are different from
the ensembles that would describe each contact taken as stand-alone device.
We further show that instead of two sets of channels with manifold cross-links
over the island the double junction can be described by pairs of channels from
both sides coherently coupled together, where each pair, however, has no coherent
connection to the others.
This finding will substantially simplify modelling transport by a Green's
functions technique. Additional channels through only one junction may complete
the picture. Finally we discuss how partial coherence across the island with
an appropriate ansatz can be modelled in the same scheme.
\end{abstract}


\section{Introduction}

In low-dimensional electron systems \cite{Grabert,Buett,imry}
the quantization of conductance \cite{beenakker} is
observed. For constrictions in two-dimensional electron gases the conductance
adopts multiples of the conductance quantum $G_0=e^2/h$ because the transverse
wave numbers determine how many modes contribute \cite{datta,quantcond}.
Single-atomic size contacts exhibit typical, reproducible and material dependent conductances,
which, however, in  general are no integer multiples of $G_0$ \cite{elke}.
The conductance
is associated to an ensemble of transport channels, the number
of which corresponds
to the number of valence orbitals of the element used. The individual channel
transmissions
in the range $0\ldots 1$ reflect the wave-function overlap from
the central to neighbouring and following atoms
\cite{cuevmicro}. Although atomistic ab initio calculations
greatly complement experiments in this field, they are not needed to deduce
transport channels and their transmissions from measurements. The description
using an ensemble of transport channels is a more general concept applicable
to any sort of point contact \cite{cuevgreen}. The contact - including leads
to bulk electrodes - can be viewed as a black box that behaves
like the associated ensemble of transport channels. A deeper interpretation
of the channels is not required. We shortly
review here, in our own way, the theoretical construction behind the description of a single
contact in terms of a channel ensemble, because this basis is needed for
generalizing the concept to the double junction.

\section{Single junction}

Regarding the left and right side of a point contact as in Fig.1a at first as
uncoupled reservoirs let
there be a finite number of orthonormal modes on each side (2 on the left
and 3 on the right as depicted in Fig.1b, for example). These may be localized atomic
orbitals or band modes of a solid crystal.
\footnote{The model later will, however,
assume that each mode has a constant density of states around the Fermi energy
in a range a few times the equivalent of the voltage applied over the contact
or the correspoding bulk superconductor quasiparticle density of states
according, for example, to BCS theory.}
(Small letters $l$ and $r$ are used here just to distinguish the single from the double
junction.) Putting the left and right side together, scattering is determined
by some complex amplitudes $s_{ij}$ gathered into matrix $S$ (Fig.1c).
The case that some modes may not couple
is included. Some entries $s_{ij}$ may be zero. Reverse scattering is given
by the adjoint matrix (Fig.1d).
Not every matrix with as many lines as there are modes on one side
and as many columns as there are modes on the other side can be a scattering
matrix $S$, though. Probability conservation sets upper limits on the
entries. A mandatory limit is, of course, that the absolute
value of each entry be lower than or equal to 1, $\vert s_{ij} \vert \le 1$ for all $i,j$. But then
one could think at first glance that it would be sufficient to demand that
a particle occupying a pure mode on one side after being scattered to the
other side should not cause a probability exceeding 1 there, that is
$\sum_i \vert s_{ij}\vert^2\le 1$ and then the other way round
$\sum_j \vert s_{ji}^*\vert^2\le 1$ for every $j$. An example showing that
this is insufficient will be given in section 4. For independent transport
channels we want an input on the left to come back into the same eigenmode
it came from there after having been scattered to the right and back.
All further multiple reflections will then stay in this mode.  
The channels will therefore
be determined by diagonalizing $S^{\dagger}S$. By construction $S^{\dagger}S$
is a hermitian matrix and thus has real eigenvalues. To ensure that the total
probability does not increase it is required that any superposition of modes
on the left described by a normalized distribution vector when scattered
to the right and back is projected onto a mode distribution with total
probability less than or at most 1 again on the left. The same, of course,
has to hold true for starting with a normalized vector on the right and
regarding the probability returning from one backreflection to the left.
We shall show that the prohibition of probability creation in forth- and
backreflections is equivalent to all eigenvalues of $S^{\dagger}S$ being less
or equal to one. ''All eigenvalues of $S^{\dagger}S$ $\le$ 1'' is not a
property to deduce of a scattering matrix $S$, but the definition to put for
calling a matrix $S$ a scattering matrix.

\begin{figure}\begin{center}
\includegraphics[width=4.5cm,angle=270]{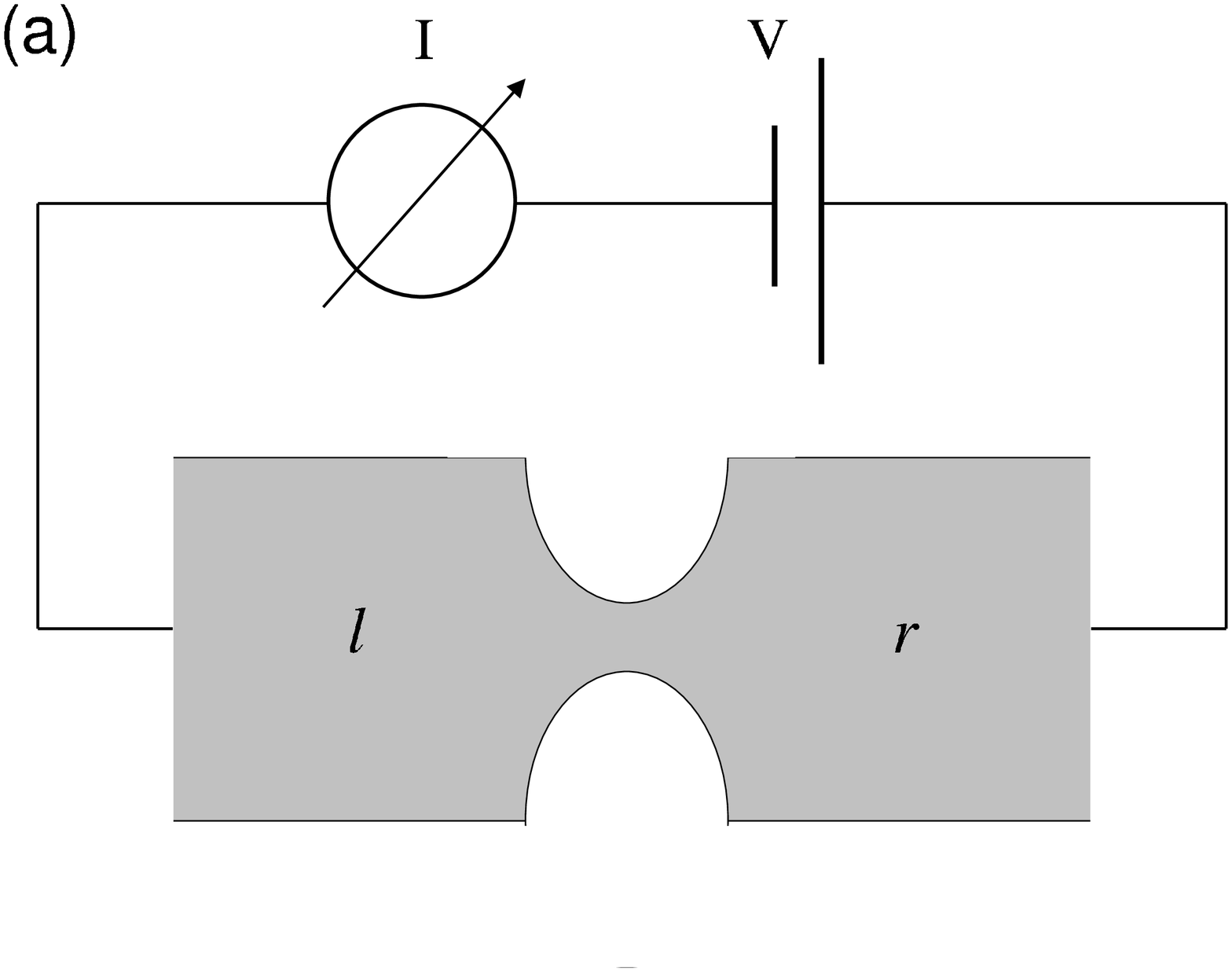}
\includegraphics[width=5cm,angle=270]{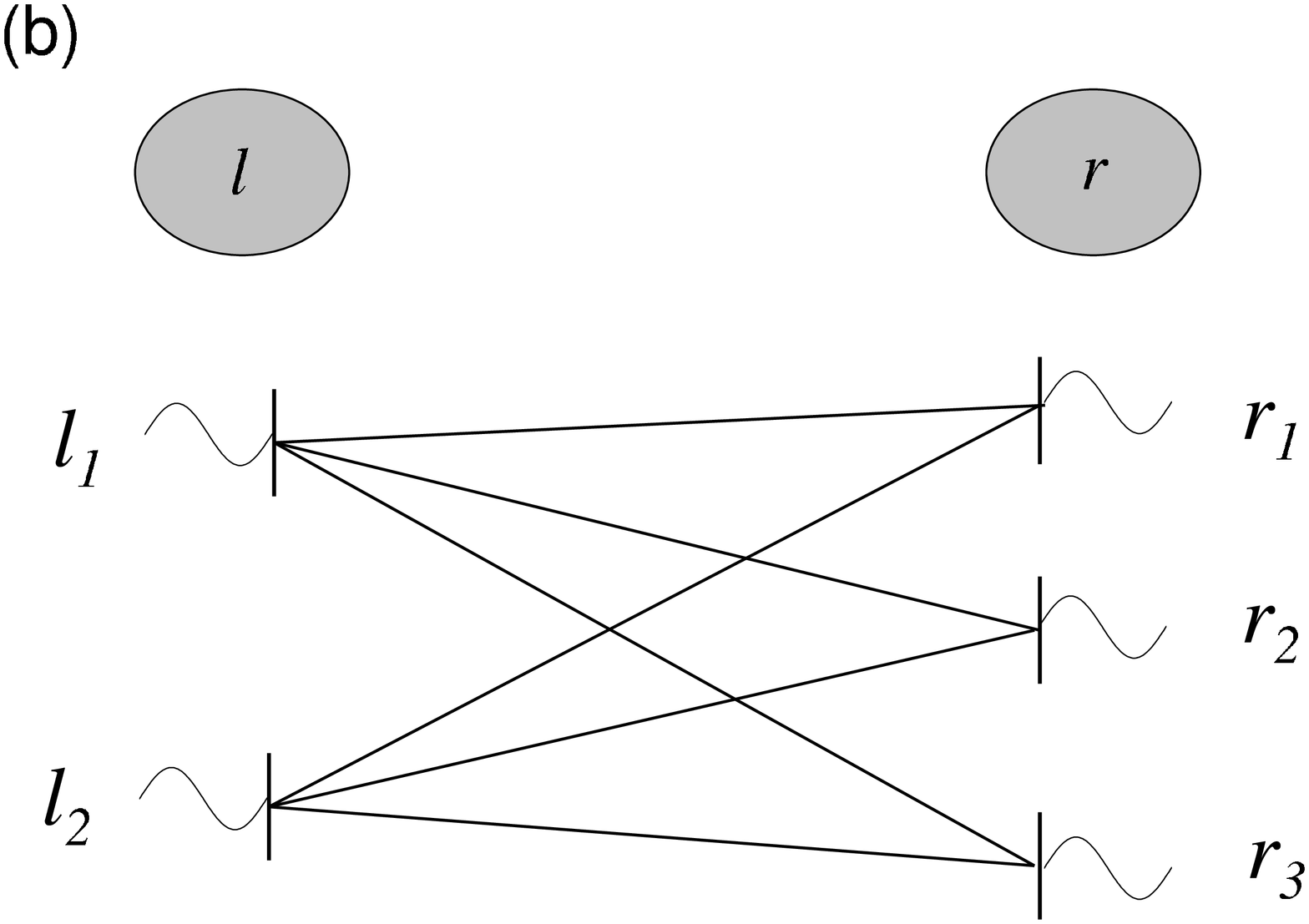}

\parbox[t]{3.75cm}{
\parbox[t]{0.25cm}{(c)}\parbox[t]{3.5cm}{
{\large
$$\offinterlineskip \tabskip 0pt \vbox{\halign{\strut
# \hfill & \vrule # & \vrule # & \hfill # \hfill & \vrule # &
\hfill # \hfill & \vrule # \cr
$S$: &&& \ $l_1$ && \ $l_2$ & \cr
\noalign{\hrule}\noalign{\hrule} 
$r_1$ & & & \ $s_{11}$ && \ $s_{12}$ & \cr
\noalign{\hrule} 
$r_2$ & & & \ $s_{21}$ && \ $s_{22}$ & \cr
\noalign{\hrule} 
$r_3$ & & & \ $s_{31}$ && \ $s_{32}$ & \cr
\noalign{\hrule} 
}} $$}}}\parbox{2.5cm}{\phantom{nix-nur-Platz}}
\parbox[t]{3.75cm}{(d) \newline
{\large
$$\offinterlineskip \tabskip 0pt \vbox{\halign{\strut
# \hfill & \vrule # & \vrule # & \hfill # \hfill & \vrule # &
\hfill # \hfill & \vrule # & \hfill # \hfill & \vrule # \cr
$S^{\dagger}$: &&& \ $r_1$ && \ $r_2$ && \ $r_3$ & \cr
\noalign{\hrule}\noalign{\hrule} 
$l_1$ & & & \ $s^*_{11}$ && \ $s^*_{21}$ && \ $s^*_{31}$ & \cr
\noalign{\hrule}
$l_2$ & & & \ $s^*_{12}$ && \ $s^*_{22}$ && \ $s^*_{32}$ & \cr
\noalign{\hrule} 
}} $$}}
\caption{(a) Point contact embedded in an electric circuit. (b) Modes left
and right that can arbitrarily couple. (c) Scattering matrix from left to
right. (d) Scattering matrix from right to left.}
\end{center}\end{figure}

In contrast to the usual scattering formalism \cite{imry,datta} we do not
distinguish incoming and outgoing modes here. Furthermore instead of two
transmission and two reflection quadrants our {\it scattering matrix} $S$ only
consists of one transmission block and only describes transmission from one
side of the contact to the other, and therefore does not have to be square.
The term {\it modes} or {\it original modes} is used for states that conveniently
describe the left and right side of the contact, which can be taken from the
situation before a contact is established. We avoid the word ''channel'' for
these, which is often used synonymous with incoming and outgoing modes \cite{
imry,datta} because it is implied differently in the experiment-related
description of atomic contacts \cite{elke}. We call {\it eigenmodes} or
{\it new modes} those linear combinations of original modes that constitute
eigenvectors of $S^{\dagger}S$ or $SS^{\dagger}$. Eigenmodes on the left and
the right side of a contact are associated to each other one to one and their
connections are called {\it transport channels} or simply {\it channels}.

Let $a_{ij}$ denote the entries of the matrix $({a\!\!\!\vert}_{S^{\dagger}S})$
such that each column of $({a\!\!\!\vert}_{S^{\dagger}S})$ is a normalized
eigenvector of $S^{\dagger}S$. Eigenvectors associated to eigenvalues zero
have to be kept, and in case of degenerate eigenvalues choose a set of
orthogonal eigenvectors associated to them.
$({a\!\!\!\vert}_{S^{\dagger}S})$ is a square matrix with the dimension the
number of modes on the left. The orthonormality of the eigenvectors is
expressed through
\begin{equation} \sum_j a_{ji}^*a_{jk}=\delta_{ik}.\end{equation}
Any normalized input state on the left given by a distribution vector $\vec{b}$
in the basis of the original modes can be converted into the basis of
eigenvectors, where we shall call this same vector $\vec{c}$.
$({a\!\!\!\vert}_{S^{\dagger}S})$ is invertable.
$\vec{b}=({a\!\!\!\vert}_{S^{\dagger}S})\vec{c}$ or $b_m=\sum_j a_{mj}c_j$.
From $\sum_m b_m^*b_m=1$ it easily follows that also $\sum_j c_j^*c_j=1$ and
vice versa.
The following calculation shows the above claimed equivalance about $S$ being
a scattering matrix:
\begin{eqnarray} && \;\; \vert S^{\dagger}S\; \vec{b}\vert^2 =
\sum_i \; \vert \sum_j (S^{\dagger}S)_{ij}b_j\vert^2 \le 1 \nonumber \\
&\Longleftrightarrow & \;\; \sum_{i,j,k} (S^{\dagger}S)_{ij}^*b_j^* \;
(S^{\dagger}S)_{ik}b_k \le 1 \nonumber \\
&\Longleftrightarrow & \;\; \sum_{i,j,k,l,m} c_l^*(S^{\dagger}S)_{ij}^*a_{jl}^* \;
(S^{\dagger}S)_{ik}a_{km}c_m \le 1 \nonumber \\
&\Longleftrightarrow & \;\; \sum_{i,l,m} c_l^*\; (\lambda_la_{il})^* \;
\lambda_ma_{im} \; c_m \le 1 \nonumber \\
&\Longleftrightarrow & \;\; \sum_m c_m^*\lambda_m^2c_m \le 1
\end{eqnarray}
$\sum_k(S^{\dagger}S)_{ik}a_{km}=\lambda_ma_{im}$ expresses that the $m$th
column of $({a\!\!\!\vert}_{S^{\dagger}S})$ is the eigenvector corresponding
to eigenvalue $\lambda_m$. Of course, $\lambda_m^*=\lambda_m$, because
eigenvalues are real, and (1)  has been used.

Some normalized vector $\vec{b}$ or equivalently any normalized
vector $\vec{c}$ can arbitrarily be chosen. $\vec{c}$ could especially
be a vector with any one
component equal to 1 and all other components 0. Then obviously
$\lambda_m^2\le 1 \Rightarrow \vert\lambda_m\vert\le 1$ for every $m$
individually.
To initialize the above calculation from the last line, suppose that
every $\vert\lambda_m\vert$ is smaller or equal to 1. Then, for a general
$\vec{c}$, (2) simply means weighing each term in the sum
$\sum_m c_m^*c_m=1$ by a factor $\lambda_m^2\le 1$,
which must give a result $\le 1$.

Having obtained the scattering amplitudes from a microscopic physical model should
ensure $S$ being a scattering matrix and all eigenvalues of $S^{\dagger}S$
less or equal to 1. However, and also because of the need to make up
number examples here, an
easier criterion than calculating all eigenvalues of $S^{\dagger}S$
would be helpful. \begin{equation}
\sum_{i,j}\vert (S^{\dagger}S)_{ij}\vert^2\le 1\end{equation}
is a sufficient condition, although not a necessary one.

As a further important aspect one may wonder whether defining transport
channels as (eigen-)modes from the right being backscattered only into
themselves would have made a difference from having required this property
for (eigen-)modes on the left. This would mean looking for the eigenvalues and
eigenvectors of $SS^{\dagger}$. As $SS^{\dagger}$ may have a different
dimension from $S^{\dagger}S$, one of these matrices may have
more eigenvalues than the other one. The conjecture, that one
might get a greater number of transport channels as well as channels with
different transmissions with an ansatz looking at backscattering from one
side instead of from the other, however, is wrong.
We shall now demonstrate that, if $SS^{\dagger}$ is of greater dimension than
$S^{\dagger}S$, $SS^{\dagger}$ has all eigenvalues that $S^{\dagger}S$ has
and all its remaining additional eigenvalues are zero.
Note again that columns of $({a\!\!\!\vert}_{S^{\dagger}S})$ are the
eigenvectors of $S^{\dagger}S$: \begin{equation}
(S^{\dagger}S)^{n_l}({a\!\!\!\vert}_{S^{\dagger}S})^{n_l}=
({a\!\!\!\vert}_{S^{\dagger}S})^{n_l}(\lambda_{S^{\dagger}S})^{n_l}
\end{equation}
$(\lambda_{S^{\dagger}S})$ is the diagonal matrix containing the eigenvalues
of $S^{\dagger}S$ in the order of the columns of $({a\!\!\!\vert}_{S^{\dagger}S})$.
To get each column vector of $({a\!\!\!\vert}_{S^{\dagger}S})$ multiplied
by the respective $\lambda_m$, $(\lambda_{S^{\dagger}S})$ has to be multiplied
to $({a\!\!\!\vert}_{S^{\dagger}S})$ from the right. $S^{\dagger}S$,
$({a\!\!\!\vert}_{S^{\dagger}S})$ and $(\lambda_{S^{\dagger}S})$ are all
square matrices with dimension the number of modes $n_l$ on the left side. Now
multiply (4) by $S$ from the left and as matrix multiplication is associative,
we can view that as \begin{equation}
(SS^{\dagger})^{n_r}S^{n_rn_l}
({a\!\!\!\vert}_{S^{\dagger}S})^{n_l}= S^{n_rn_l}
({a\!\!\!\vert}_{S^{\dagger}S})^{n_l}(\lambda_{S^{\dagger}S})^{n_l}
\end{equation}
(5) tells us that the columns of $S({a\!\!\!\vert}_{S^{\dagger}S})$ are
eigenvectors of $SS^{\dagger}$ - though not necessarily normalized - with
the entries of $(\lambda_{S^{\dagger}S})$ as associated eigenvalues. $SS^{\dagger}$ is a
square matrix with dimension the number of modes $n_r$ on the right side. $S$
as well as $S({a\!\!\!\vert}_{S^{\dagger}S})$ has $n_r$ lines and $n_l$
columns. Single upper indices in (4) and (5) denote the dimension
of a square matrix, double upper indices the line and column number of a
rectangular matrix. Every eigenvalue $\lambda_m$ of $S^{\dagger}S$ is an
eigenvalue of $SS^{\dagger}$, too. However, we have not yet specified which
number of modes, $n_l$ or $n_r$, is the greater. For equal mode numbers $n_l=n_r$
we have just proven that the sets of eigenvalues of $S^{\dagger}S$ and
$SS^{\dagger}$ are identical. Now firstly suppose that $n_l<n_r$. Then (5)
is a statement about $n_l$ out of the $n_r$ eigenvalues and eigenvectors
of $SS^{\dagger}$, and gives no information about the other $n_r$-$n_l$. Secondly suppose
that $n_l>n_r$. As a preparation for the following argument we show
that the scalar product of any two different columns of
$S({a\!\!\!\vert}_{S^{\dagger}S})$ vanishes. Elements of (4) multiplied by
$({a\!\!\! -}^*_{S^{\dagger}S})$ from the left give exactly those scalar
products. $({a\!\!\! -}^*_{S^{\dagger}S})$ is the matrix with the complex
conjugates of the eigenvectors of $S^{\dagger}S$ as lines, or
$({a\!\!\!\vert}_{S^{\dagger}S})^{\dagger}$. $a_{ij}$ as before refers to the
entries of $({a\!\!\!\vert}_{S^{\dagger}S})$.
\begin{eqnarray}
\sum_{i,j,m}S_{ij}a_{jk}(S_{im}a_{ml})^*=\sum_{i,j,m}a_{ml}^*S_{im}^*S_{ij}a_{jk}=
\nonumber \\
\sum_{i,j,m}a_{ml}^*S^{\dagger}_{mi}S_{ij}a_{jk}=\sum_ma_{ml}^*\lambda_ka_{mk}
=\delta_{lk}\lambda_k
\end{eqnarray}
$S({a\!\!\!\vert}_{S^{\dagger}S})$ has $n_l$ columns, each of which is an
$n_r$-vector. In an $n_r$-dimensional vector space there can, however, only
be $n_r$ non-vanishing mutually orthogonal vectors. To fulfill (6) the
remaining $n_l$-$n_r$ column vectors must be the zero vector. With more modes
on the left than the right there are necessarily some modes or linear
combinations of modes on the left that do not get transmitted through the junction at all.
Now in (6) choose $k=l$ and $k$ one of those column numbers for which
$\sum_jS_{ij}a_{jk}$ is zero for all $i$. Then it follows that the
corresponding $\lambda_k$, which is an eigenvalue of $S^{\dagger}S$, is
zero as well. For the single junction diagonalizing $S^{\dagger}S$ or
$SS^{\dagger}$ will lead to the same ensemble of transport channels. For
illustration an example with numbers is given in section 4.
Regarding the first and last expression in the equation for $l=k$,
(6) also tells us that each $\lambda_k$ represents the sum of some absolute values
squared. Therefore by its special contruction $S^{\dagger}S$ not only is
a hermitian matrix with real eigenvalues, but all eigenvalues are even
greater or equal to zero. This is essential because we shall identify the
square roots of the $\lambda_m$ as phaseless transmission amplitudes $t$.

\section{Double junction}

Properties of a series of two junctions will initially be given in terms of
two scattering matrices, one between reservoirs L(left) and I(island), and
one between reservoirs I and R(right) (Fig.3a).\footnote{
Still the respective reverse scattering matrices are the complex conjugate
transposed ones. With an extended island one could think of a charge being
backscattered immediately at the island edge or crossing the island, being
backscattered at the rear edge and thus resonator-like interference. For
coherent coupling between junctions we are thinking of the island longitudinal
dimension between the contacts comparable to the electron coherence length.
The transverse dimension may be larger. In contrast to a quantum dot with
discrete levels, we want to regard a bulk-like island. The set of original
modes on the island should be the same in both scattering matrices, and
therefore consist of eigenstates of this bounded space. We thus set the
premises for still having a continuous and quasi-constant DOS as a function
of energy, however, no extra phase factors to scattering amplitudes
complicating the calculation for charges going across the island.}
Multiplying each scattering matrix by its adjoint and diagonalizing would
give transport channel ensembles for each junction separately and represent
the setup for incoherent coupling \cite{my06} across the island. There are
multiple reflections in each junction, but a charge transported to the island
not to go back through that same junction
relaxes there, that is changes the island's electrostatic potential,
however, looses the information which (eigen-)mode it had come into. Diagonalizing
matrices for each junction separately will not lead to the same linear
combinations of island modes as eigenmodes for each side.
Setting up our formalism to determine transport channels with coherent coupling
between two the junctions is not obvious, should just any of the three reservoirs
L,I,R be chosen to demand that eigenmodes there associated to channels exclusively
return to themselves if transmitted away from that reservoir and backscattered
into it. Nevertheless, as a scattering process starting on the island after
two scattering steps offers no other possibility than to have brought the charge
carrier back to the island, the above requirement {\it for the island} provides
a promising ansatz.

\begin{figure}
\parbox[t]{8cm}{(a)
$$\offinterlineskip \tabskip 0pt \vbox{\halign{\strut
# \hfill & \vrule # & \vrule # &
\hfill # \hfill & \vrule # &
\hfill # \hfill & \vrule # &
\hfill # \hfill & \vrule # &
\hfill # \hfill & \vrule # \cr
$S$: &&& \ $I_1$ && \ $I_2$ && \ $I_3$ && \ $I_4$ & \cr
\noalign{\hrule}\noalign{\hrule}
\ $L_1$ &&& \ $S_{L1I1}$ && \ $S_{L1I2}$ && \ $S_{L1I3}$ && \ $S_{L1I4}$ & \cr
\noalign{\hrule}
\ $L_2$ &&& \ $S_{L2I1}$ && \ $S_{L2I2}$ && \ $S_{L2I3}$ && \ $S_{L2I4}$ & \cr
\noalign{\hrule}
\ $L_3$ &&& \ $S_{L3I1}$ && \ $S_{L3I2}$ && \ $S_{L3I3}$ && \ $S_{L3I4}$ & \cr
\noalign{\hrule}\noalign{\hrule}
\ $R_1$ &&& \ $S_{R1I1}$ && \ $S_{R1I2}$ && \ $S_{R1I3}$ && \ $S_{R1I4}$ & \cr
\noalign{\hrule}
\ $R_2$ &&& \ $S_{R2I1}$ && \ $S_{R2I2}$ && \ $S_{R2I3}$ && \ $S_{R2I4}$ & \cr
\noalign{\hrule}
}} $$}
\parbox{1cm}{\phantom{nix}}
\parbox[t]{8cm}{(b)
$$\offinterlineskip \tabskip 0pt \vbox{\halign{\strut
# \hfill & \vrule # & \vrule # &
\hfill # \hfill & \vrule # &
\hfill # \hfill & \vrule # &
\hfill # \hfill & \vrule # & \vrule # &
\hfill # \hfill & \vrule # &
\hfill # \hfill & \vrule # \cr
$S^{\dagger}$: &&& \ $L_1$ && \ $L_2$ && \ $L_3$ &&& \ $R_1$ && \ $R_2$ & \cr
\noalign{\hrule}\noalign{\hrule}
\ $I_1$ &&& \ $S^*_{L1I1}$ && \ $S^*_{L2I1}$ && \ $S^*_{L3I1}$ &&& \ $S^*_{R1I1}$ && \ $S^*_{R2I1}$ & \cr
\noalign{\hrule}
\ $I_2$ &&& \ $S^*_{L1I2}$ && \ $S^*_{L2I2}$ && \ $S^*_{L3I2}$ &&& \ $S^*_{R1I2}$ && \ $S^*_{R2I2}$ & \cr
\noalign{\hrule}
\ $I_3$ &&& \ $S^*_{L1I3}$ && \ $S^*_{L2I3}$ && \ $S^*_{L3I3}$ &&& \ $S^*_{R1I3}$ && \ $S^*_{R2I3}$ & \cr
\noalign{\hrule}
\ $I_4$ &&& \ $S^*_{L1I4}$ && \ $S^*_{L2I4}$ && \ $S^*_{L3I4}$ &&& \ $S^*_{R1I4}$ && \ $S^*_{R2I4}$ & \cr
\noalign{\hrule}
}} $$}

\caption{(a) Scattering matrix and (b) adjoint for the double junction.
}\end{figure}

\begin{figure}
\includegraphics[width=6cm,angle=270]{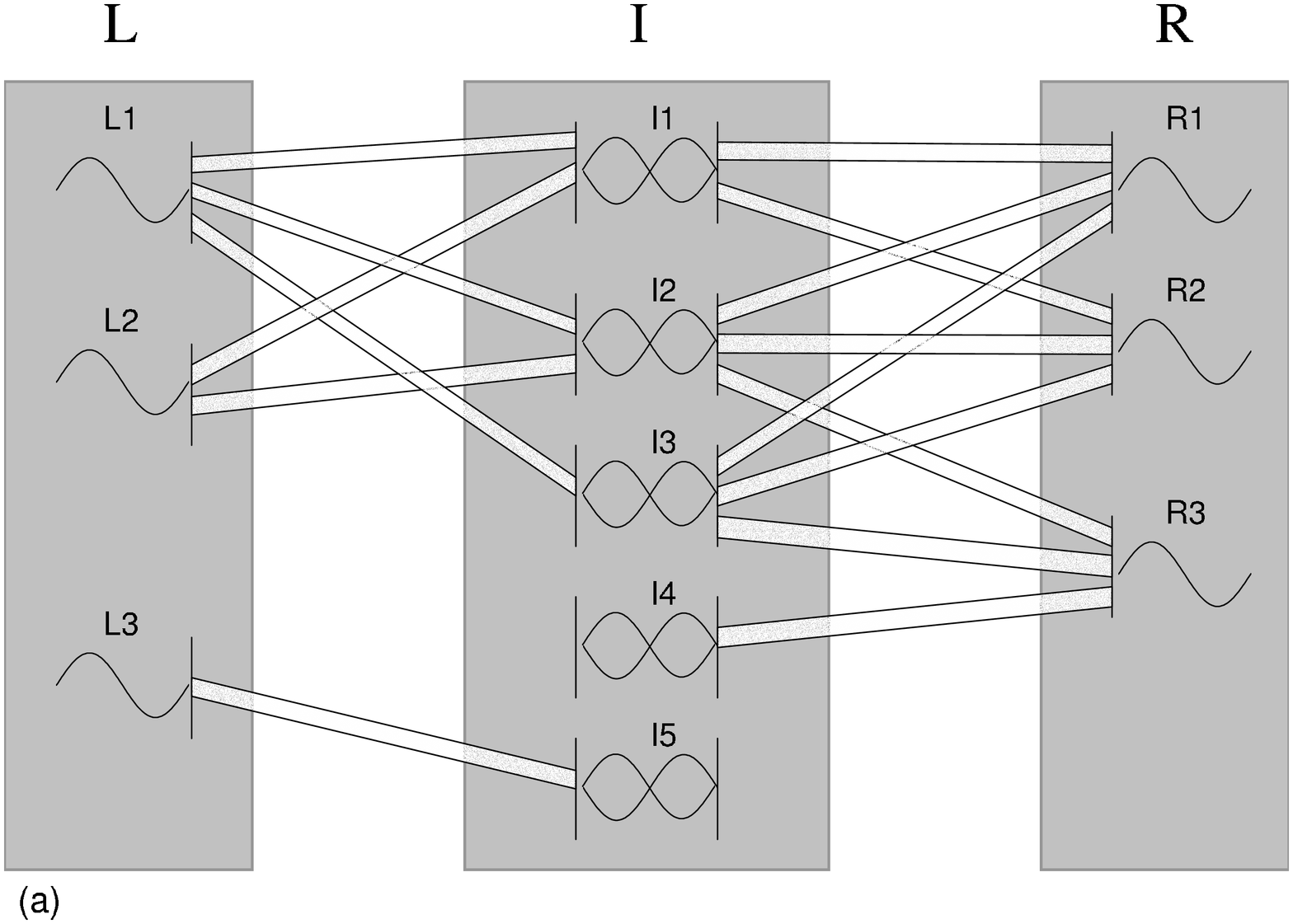}\phantom{nix}
\includegraphics[width=6cm,angle=270]{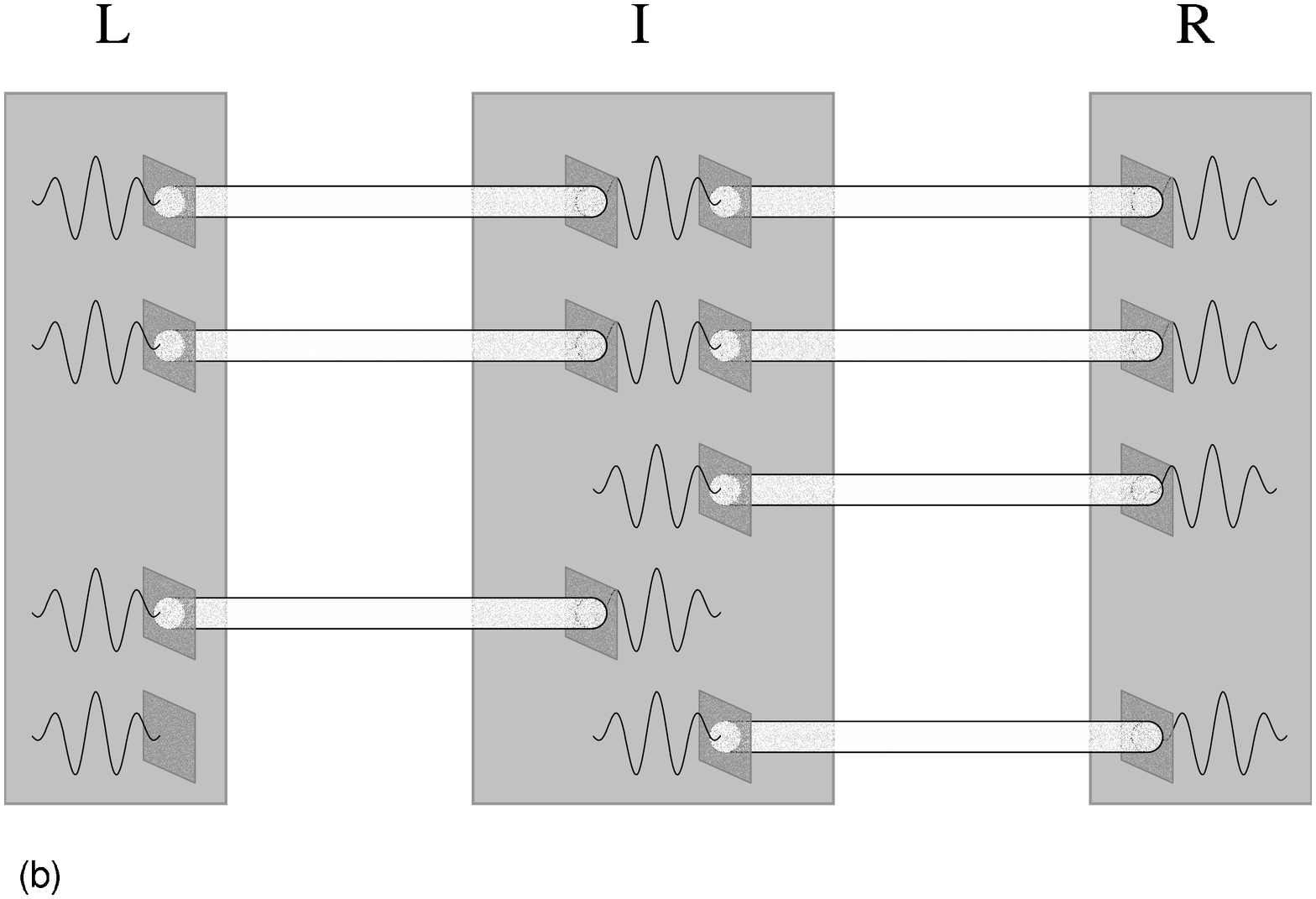}
\caption{(a) Connections of original modes in a left and right
reservoir and on an island between them as given by scattering matrices.
(b) Eigenmodes as connected by transport channels.}
\end{figure}

Therefore given scattering matrices from I to L and from I to R, with
by definition the same number of I-modes, are just joint together into one
scattering matrix $S$ by taking all rows together. $S$ and $S^{\dagger}$
are formally given with exemplary numbers of modes three on the left,
four on the island and two on the right in Fig.2 (Fig.3 are drawn for
different numbers). Now $S^{\dagger}S$ is diagonalized, again denoting
eigenvalues as $\lambda_m$ and eigenvectors as $({a\!\!\!\vert}_{S^{\dagger}S})$.
The column vectors from $({a\!\!\!\vert}_{S^{\dagger}S})$ build linear combinations of the original
island modes. Eigenvectors of $SS^{\dagger}$, however, given by
$S({a\!\!\!\vert}_{S^{\dagger}S})$, now combine modes from both leads L and R.
A $\sqrt{\lambda_m}$ associated to an eigenvector out of $({a\!\!\!\vert}_{S^{\dagger}S})$
or the corresponding vector from $S({a\!\!\!\vert}_{S^{\dagger}S})$ gives us the
transmission amplitude of a ''combined channel'' between the island and the
entire outside world consisting of both leads. If there have been $n_L$
original modes at L and $n_R$ at R, every column of $\frac{1}{\sqrt{\lambda_m}} S
({a\!\!\!\vert}_{S^{\dagger}S})$ is a normalized vector of length $n_L$+$n_R$,
whose first $n_L$ components represent coefficients for L-modes and whose
last $n_R$ components are coefficients for R-modes. However, like
for the single junction case, modes at L and R are independent.\footnote{
With the single junction we usually do not regard the mapping of the original
modes on a side onto the eigenmodes for transmission any more. For the supply
of charge carriers the original modes are considered to be incoherently
populated with probability one up the Fermi energy, which automatically makes
the eigenmodes be populated each with probability one on each side for they
are normalized linear combinations.}
Supplying a charge carrier in an original mode on the left, for example,
makes the transmission eigenmodes it contributes to be populated with
percentages the absolute values squared of their coefficients in the respective
column from $\frac{1}{
\sqrt{\lambda_m}}S({a\!\!\!\vert}_{S^{\dagger}S})$. Therefore the transmission
of an above mentioned combined channel splits up into the probability
percentages left and right side modes contribute to the respective eigenmode.
For assigning transmission amplitudes we take the roots of these parts to be
multiplied by $\sqrt{\lambda_m}$. So for the transmission amplitudes $t_L$
and $t_R$ for the left and right side channels of a pair of channels coherently
linked together on the island (=''combined channel'') we obtain
\begin{equation}
t_L=\sqrt{\lambda_m}\sqrt{\sum_{i=1}^{n_L}\vert \frac{1}{\sqrt{\lambda_m}}
(S({a\!\!\!\vert}_{S^{\dagger}S}))_{im}\vert^2}
\end{equation}
and
\begin{equation}
t_R=\sqrt{\lambda_m}\sqrt{\sum_{i=n_L+1}^{n_L+n_R}\vert \frac{1}{\sqrt{\lambda_m}}
(S({a\!\!\!\vert}_{S^{\dagger}S}))_{im}\vert^2}
\end{equation}
This construction demonstrates that a series of point contacts enclosing a
bulk-like island even with coherent coupling across the latter is {\it not}
equivalent simply to a number of effective transport channels from left to right.
Paths from left to right rest a channel through the left junction connected to a
channel through the right junction and these two parts can have different
transmissions. However, we cannot expect to describe coherently coupled
junctions by the channel ensembles each one would exhibit as a single junction.
Furthermore, by construction there is no coherent cross coupling on the island
between our newly found channels. One channel from the left can only be coherently
coupled to one channel from the right and vice versa (Fig.3b). This finding
elegantly solves the problem of generalizing a Green's functions method for
modelling current-voltage characteristics in the coherent case \cite{my07}
to several channels per junction. The algorithm to calculate changing rates
for the island charge is only needed for a single channel per junction.
Contributions from all pairs of channels can then simply be added in classical
rate equations. There may also be channels in a junction not coherently coupled
to a partner channel in the other junction (or pairs with transmission amplitude
zero in one half). Their island charging rates are even easier to calculate
\cite{my06}. We should only expect such unpaired channels, however, if in the
original pattern there are modes coupled across one junction that are totally
decoupled from all others involved in a network of couplings over both
junctions, like L3 and I5 in Fig.3a. I4 is not directly
coupled to L, however indirectly via R3 and I3 as well as other paths, and
will thus contribute to eigenmodes involving connections to both leads. Like
in the single junction case, there may be specific linear combinations of
original modes at every one of the three sites that are not transmitted at all.

For a consistent description of the double junction with coherent coupling
across the island, it has to be required that $\vert S^{\dagger}S\vec{b}\vert
\le 1$ with the normalized vector $\vec{b}$ representing any linear combination
of original modes on the island and $S$ and $S^{\dagger}$ now the big matrices
from Fig.2. The prohibition of probability creation here also implies that all
eigenvalues $\lambda_m$ of $S^{\dagger}S$ have to be less than or at most one,
even if the transmissions assigned to the channels in each junction further
get multiplied with the parts by which a new combined island mode is transferred
to each side. The new left and right modes at the ends of a pair of channels
in fact constitute only parts of new combined-lead modes.
The latter, however, form a unique set of eigenmodes.
All these considerations about the double junction will be illustrated
by number examples in the next section.

\begin{figure}
\phantom{Zeile}

\vspace{1cm}

\parbox{8cm}{
\includegraphics[width=6.5cm]{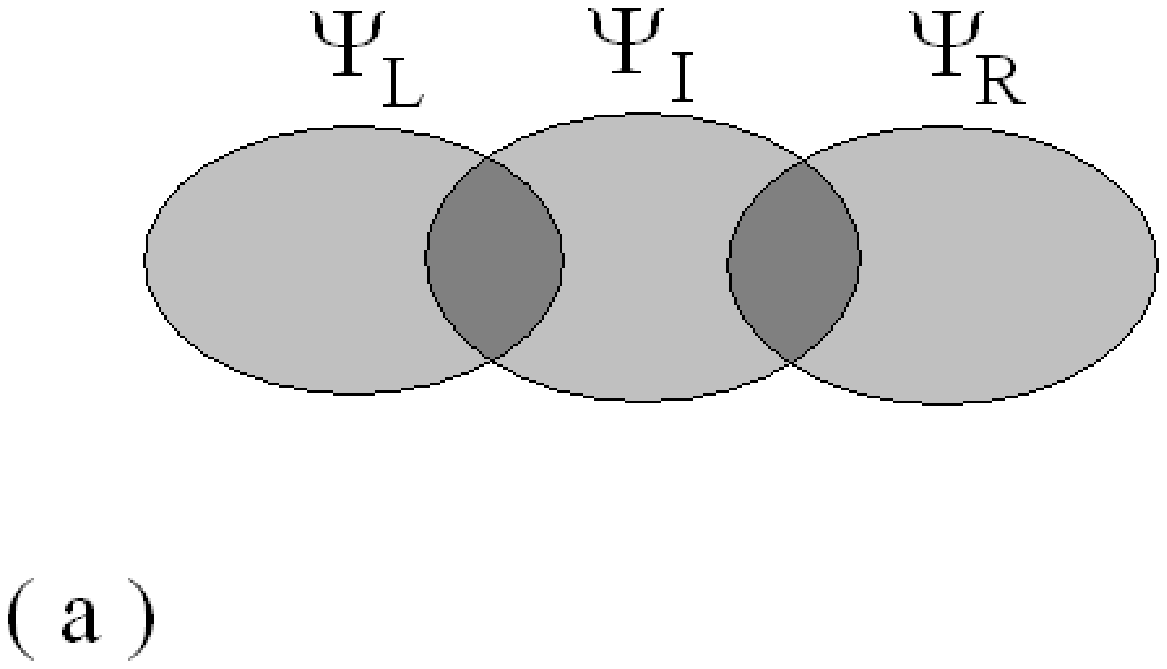}}
\parbox{1cm}{\phantom{nix}}
\parbox{8cm}{
\includegraphics[width=6.5cm]{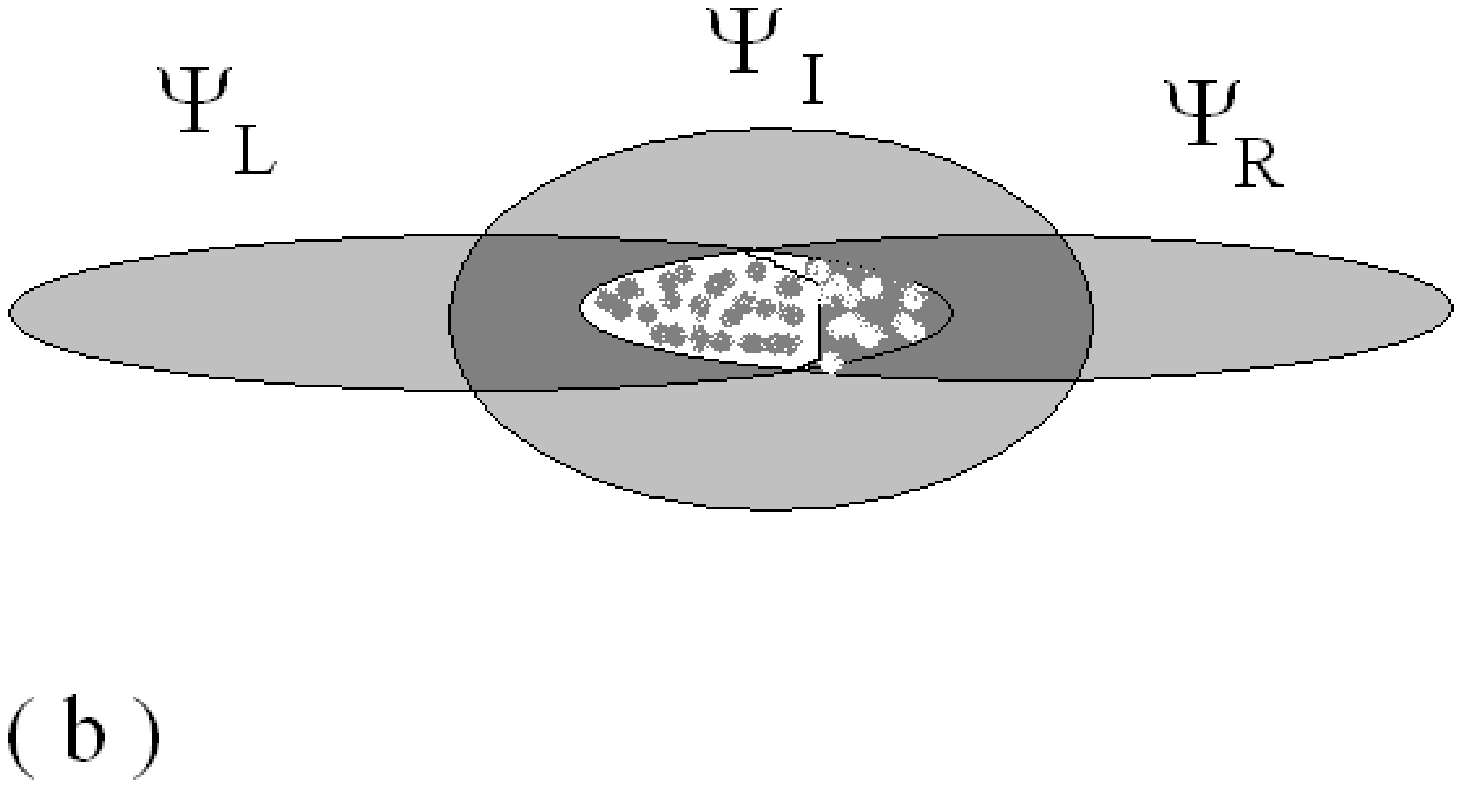}}

\parbox{8cm}{
\includegraphics[width=4.5cm,angle=270]{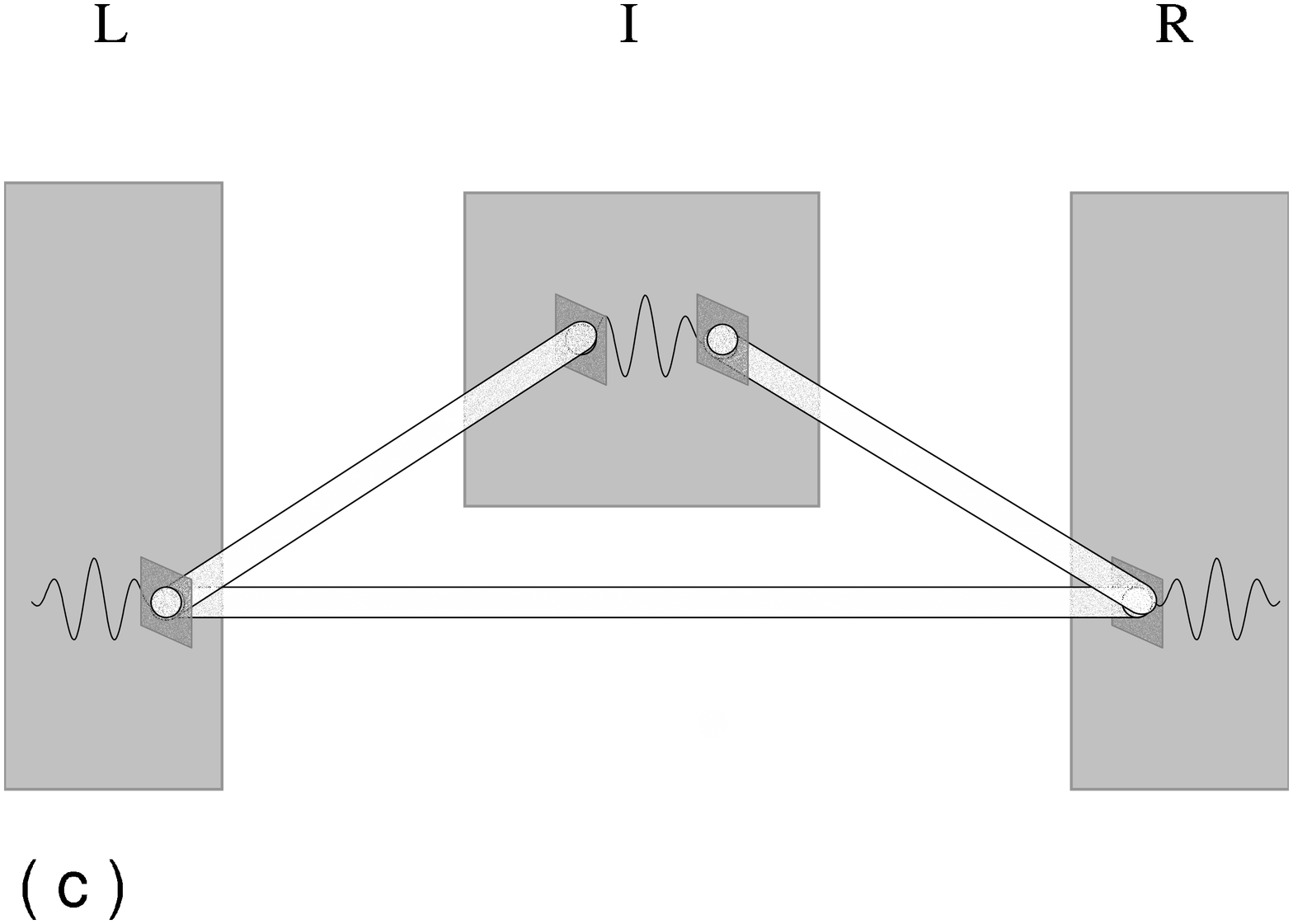}}
\parbox{1cm}{\phantom{nix}}
\parbox{8cm}{\small {\bf Figure 4.}
(a) Overlaps of $\Psi_I$ with $\Psi_L$ and $\Psi_R$ well separated.
(b) In case of part of $\Psi_I$ overlapping with both $\Psi_L$ and $\Psi_R$
this can in an auxiliary way be supposed to scatter to L or R only with certain
weight factors. (c) Our model does not forsee direct coupling between L and R
leading to direct channels in parallel with paths via the
island.}
\end{figure}

\setcounter{figure}{4}

We shall here present some further more marginal considerations on probability
conservation or prohibiting probability creation. At the beginning of our
analysis we regarded the system as described by an orthogonal set of modes
on the left, an orthogonal set on the island and an orthogonal set on the
right. Of course, with the contact established overlaps exist between modes
in a lead and modes on the island. These determine the transition amplitudes
put into the original scattering matrix.
The overlaps should, however, only go so far that parts of an
island wave function that couple to left wave functions rest well separated
from parts that couple to the right. Then the sum of absolute squares of all
scattering matrix elements out of an island mode will in no case exceed unity.
The orthogonality of the modes in one lead ensures that the sum of absolute
squares of all couplings to one certain island mode is limited by the norm
one of that mode. Taking just any overlaps that could have been set up as
single junction scattering amplitudes does
not automatically guarantee that the norm of any column vector of the big
matrix $S$, which contains scattering to the left {\it and} right modes,
is less or equal to one. If the intersection of
the overlaps of an
island wave function with left wave functions and that with right wave functions
is not empty, there would be a direct overlap between left and right wave
functions, which will give rise to a direct transport channel from left to
right. Such a situation could be
realizable, but our model does not cover that.
Nevertheless, suppose that two junctions in series have been investigated with
incoherent coupling \cite{my06}. In the incoherent model changes of the island
charge and thus potential are felt by both contacts, but otherwise the latter
are independent and behave like a single junction each. With incoherent
coupling the regarded set of island modes need not even be the same concerning
transport through the left or the right junction, but here suppose it is.
To compare with the incoherent behaviour, one might want to estimate the
system's properties if the same junctions get linked together coherently
(without allowing for direct left-to-right channels) and
therefore just put together two single-junction scattering matrices $S_{LI}$
and $S_{RI}$ into one like in Fig.2a. Where this should result in a column
norm of $S$ exceeding unity, in this situation it makes sense to renormalize
scattering amplitudes in the way that of the island mode the part overlapping
with both left and right wave functions is scattered to either side with
probabilities proportional to the ratio of absolute squares of the single-junction
scattering amplitudes. In any case put-together scattering matrices $S$ should
be checked for no column norm exceeding unity prior to solving for transport
channels.

\section{Examples with numbers}

Let us illustrate that the simple criterion of norms of lines or columns
being less or equal to 1 is not sufficient to ensure that a matrix is a
scattering matrix. For simplicity an example with purely real numbers is
given. For the single junction
let $S^{\dagger}$ be
\begin{equation}
S^{\dagger}=\pmatrix{ 0.3 & 0.5 & 0.5 & 0.6 \cr
0.1 & 0.4 & 0.8 & 0.2 \cr}\end{equation}
supposing two modes on the left and four on the right. Then
\begin{eqnarray} &&
0.3^2+0.5^2+0.5^2+0.6^2=0.95,0.1^2+0.4^2+0.8^2+0.2^2=0.85,\nonumber \\ &&
0.3^2+0.1^2=0.10, \; 0.5^2+0.4^2=0.41, \nonumber \\ &&
0.5^2+0.8^2=0.89, \; 0.6^2+0.2^2=0.40 \end{eqnarray}
are all less than 1.
\begin{equation} S^{\dagger}S=
\pmatrix{ 0.3 & 0.5 & 0.5 & 0.6 \cr 0.1 & 0.4 & 0.8 & 0.2 \cr }
\pmatrix{ 0.3 & 0.1 \cr 0.5 & 0.4 \cr 0.5 & 0.8 \cr 0.6 & 0.2 \cr
}=\pmatrix{ 0.95 & 0.75 \cr 0.75 & 0.85 \cr }
\end{equation}
The eigenvalues of this matrix, namely the solutions of
\begin{equation}
(0.95-\lambda)(0.85-\lambda)-0.75\cdot 0.75=0,
\end{equation}
however, are $\lambda_1$=0.148 and $\lambda_2$=1.652 (rounded to 3 digits),
of which $\lambda_2$ clearly is greater than 1. (3) is not fulfilled:
\begin{equation}
0.95^2+0.75^2+0.75^2+0.85^2=2.75>1 
\end{equation}

Let us now regard
\begin{equation}
S=\pmatrix{ 0.15+0i & 0-0.50i \cr
0.10-0.15i & 0.10-0.05i \cr 0.15-0.30i & 0.05+0i\cr },
\end{equation}
which is a scattering matrix, but otherwise its complex entries are
randomly chosen. As $S$ is 3x2, there are two original modes on the left
and three on the right. \begin{equation} S^{\dagger}S
=\pmatrix{ 0.1675 & 0.025-0.05i \cr
0.025+0.05i & 0.265 } \end{equation}
and the eigenvalues of this 2x2 square matrix are $\lambda_1$=0.1421 and
$\lambda_2$=0.2904. \begin{equation} SS^{\dagger}=
\pmatrix{ 0.2725 & 0.04-0.0275i & 0.0225+0.02i \cr
0.04+0.0275i & 0.045 & 0.065+0.005i \cr
0.0225-0.02i & 0.065-0.005i & 0.115 \cr },
\end{equation}
is a 3x3 square matrix with eigenvalues $\lambda_1$=0.1421, $\lambda_2$=0.2904
and $\lambda_3$=0 (a numerical calculation gives $2.7 \cdot 10^{-17}$).
To check (3) for both $S^{\dagger}S$ and $SS^{\dagger}$ we calculate
\begin{eqnarray} 
0.1675^2+2\cdot\vert 0.025-0.05i\vert^2+0.265^2 = \nonumber\\
0.2725^2+0.045^2+0.115^2+
2\cdot\vert 0.04-0.0275i\vert^2+2\cdot\vert 0.0225+0.02i\vert^2 \nonumber \\
+\; 2\cdot\vert 0.065+0.005i\vert^2 = 0.105<1 \end{eqnarray}
By the way, the same value is obtained as
\begin{equation}
0.2904^2+0.1421^2=0.105
\end{equation}
With complex numbers even normalized eigenvectors are only determined up to a phase common to
all components. The numerical routine we used to evaluate them gave them with
real last component. For $S^{\dagger}S$ they are \begin{equation}
\pmatrix{ 0.407-0.814i\cr -0.414+0i\cr } \quad {\rm and} \quad
\pmatrix{ 0.185-0.370i\cr 0.910+0i\cr } \end{equation}
associated to $\lambda_1$ and $\lambda_2$, respectively.
Eigenvectors of $SS^{\dagger}$ are \begin{equation}
\pmatrix{ -0.085+0.123i\cr 0.859-0.003i\cr -0.490+0i\cr },\quad 
\pmatrix{ 0.277+0.020i\cr -0.457+0.043i\cr -0.844+0i\cr },\quad
\pmatrix{ 0.881+0.354i\cr 0.162+0.161i\cr 0.218+0i } \end{equation}
associated to eigenvalues zero, $\lambda_1$ and $\lambda_2$ in that order.
By multiplying (5) by $({a\!\!\! -}^*_{S^{\dagger}S})S^{\dagger}$ from the left,
one deduces that the factor needed to normalize a column vector of
$S({a\!\!\!\vert}_{S^{\dagger}S})$, which we showed to be an eigenvector of
$SS^{\dagger}$, is $1/\sqrt{\lambda_m}$ with $\lambda_m$ the corresponding
eigenvalue. We shall now check explicitly the relation between the two sets
of eigenvectors in our example, and especially verify that for unequal mode
numbers left and right, the transformation in one direction produces some
zero vector.

\centerline{\tiny \phantom{Leerzeile}}

\centerline{$ {\displaystyle
\frac{1}{\sqrt{0.1421}} \pmatrix{ 0.15+0i & 0-0.5i\cr 0.1-0.15i & 0.1-0.05i\cr
0.15-0.3i & 0.05+0i\cr}\pmatrix{ 0.407-0.814i\cr -0.414+0i\cr}=e^{0.875i}
\pmatrix{0.277+0.02i\cr -0.457+0.043i\cr -0.844+0i\cr} }$}

\centerline{\tiny \phantom{Leerzeile}}

\centerline{$ {\displaystyle \frac{1}{\sqrt{0.2904}}
\pmatrix{ 0.15+0i & 0-0.5i\cr 0.1-0.15i & 0.1-0.05i\cr 0.15-0.3i & 0.05+0i\cr
}\pmatrix{ 0.185-0.37i\cr 0.91+0i\cr}=
e^{-1.9i}\pmatrix{ 0.881+0.354i\cr
0.162+0.161i\cr 0.218\cr } }$}
\begin{equation} \end{equation}

\centerline{\tiny \phantom{Leerzeile}}

\centerline{$ {\displaystyle \frac{
\pmatrix{ 0.15+0i & 0.1+0.15i & 0.15+0.3i\cr 0+0.5i & 0.1+0.005i & 0.05+0i\cr}
}{\sqrt{0.1421}}
\pmatrix{ 0.277+0.02i\cr -0.457+0.043i\cr -0.844+0i\cr}= e^{-0.875i}
\pmatrix{ 0.407-0.814i\cr -0.414+0i\cr} }$}

\centerline{\tiny \phantom{Leerzeile}}

\centerline{$ {\displaystyle \frac{
\pmatrix{ 0.15+0i & 0.1+0.15i & 0.15+0.3i\cr
0+0.5i & 0.1+0.005i & 0.05+0i\cr }}{\sqrt{0.2904}}
\pmatrix{ 0.881+0.354i\cr 0.162+0.161i\cr 0.218+0i\cr} 
= e^{1.9i}\pmatrix{ 0.185-0.37i\cr 0.91+0i\cr} }$}

\centerline{\tiny \phantom{Leerzeile}}

\centerline{$ {\displaystyle
\pmatrix{ 0.15-0i & 0.1+0.15i & 0.15+0.3i\cr 0+0.5i & 0.1+0.005i & 0.05-0i\cr
}\pmatrix{ -0.085+0.123i\cr 0.859-0.003i\cr -0.49+0i\cr
}=
\pmatrix{ 1\cdot 10^{-4}+0i\cr 5\cdot
10^{-5}+1.5\cdot 10^{-4}i\cr} }$}
\begin{equation} \end{equation}

The last is the zero vector within the accuracy taking into account no more
than the written digits.
It cannot be normalized, of course, hence there is
no factor $1/\sqrt{\lambda_m}$ in the last line.

Entries of scattering matrices are in general complex, but for simplicity
we give examples with real numbers for the double junction, which is
totally sufficient to show the important aspects of the calculation.
To better keep track of which scattering amplitudes belong to the left and
the right junction, $S$ is noted in table form like in Fig.2. Our first
example has four modes on the island, two in each lead and no vanishing
scattering amplitudes.

\centerline{\tiny \phantom{Leerzeile}}

\begin{equation} \offinterlineskip \tabskip 0pt \vbox{\halign{\strut
# \hfill & \vrule # & \vrule # &
\hfill # \hfill & \vrule # &
\hfill # \hfill & \vrule # &
\hfill # \hfill & \vrule # &
\hfill # \hfill & \vrule # \cr
$S$: &&& \ $I_1$ && \ $I_2$ && \ $I_3$ && \ $I_4$ & \cr
\noalign{\hrule}\noalign{\hrule}
\ $L_1$ &&& \ 0.17 && \ 0.28 && \ 0.39 && \ 0.06 & \cr
\noalign{\hrule}
\ $L_2$ &&& \ 0.06 && \ 0.22 && \ 0.40 && \ 0.11 & \cr
\noalign{\hrule}
\ $R_1$ &&& \ 0.28 && \ 0.11 && \ 0.33 && \ 0.39 & \cr
\noalign{\hrule}
\ $R_2$ &&& \ 0.06 && \ 0.33 && \ 0.11 && \ 0.17 & \cr
\noalign{\hrule}
}} \end{equation}
The sum of squares of all these 16 elements is 0.9921, which is less than 1
and should thus ensure that all eigenvalues will be less than or at most 1.
The matrix to diagonalize is \begin{equation}
S^{\dagger}S=\pmatrix{
0.1145 & \ 0.1114 & \ 0.1893 & \ 0.1362 \cr
0.1114 & \ 0.2478 & \ 0.2698 & \ 0.1400 \cr
0.1893 & \ 0.2698 & \ 0.4331 & \ 0.2148 \cr
0.1362 & \ 0.1400 & \ 0.2148 & \ 0.1967 \cr
}\nonumber\end{equation}

As eigenvalues and associated eigenvectors one finds

\centerline{\tiny \phantom{Leerzeile}}

\parbox{12cm}{
{\small \begin{tabular}{ccccc}
$\lambda_m$: & 0.0080 & 0.0518 & 0.0959 & 0.8364 \\
{\tiny \phantom{nix}} & & & &  \\
$\matrix{ I_1: \cr I_2: \cr  I_3: \cr I_4: \cr }$ &
$\pmatrix{ 0.8549 \cr 0.1327 \cr -0.2433 \cr -0.4386 \cr }$ &
$\pmatrix{ 0.0728 \cr -0.6389 \cr 0.6469 \cr -0.4100 \cr }$ &
$\pmatrix{ -0.3895 \cr 0.5854 \cr 0.1872 \cr -0.6859 \cr }$ &
$\pmatrix{ 0.3349 \cr 0.4812 \cr 0.6981 \cr 0.4111 \cr }$ 
\end{tabular}}}\parbox{2.5cm}{
\begin{equation} \end{equation}}

\centerline{\tiny \phantom{Leerzeile}}

$S$ times the matrix made of these four column vectors returns a matrix
consisting of the following four column vectors:

\centerline{\tiny \phantom{Leerzeile}}

\parbox{12cm}{
{\small \begin{tabular}{ccccc}
$\matrix{ L_1: \cr L_2: \cr R_1: \cr R_2: \cr }$ &
$\pmatrix{ 0.0627 \cr -0.0651 \cr 0.0026 \cr -0.0062 \cr }$ &
$\pmatrix{ 0.0612 \cr 0.0775 \cr 0.0039 \cr -0.2050 \cr }$ &
$\pmatrix{ 0.1296 \cr 0.1048 \cr -0.2504 \cr 0.0738 \cr }$ &
$\pmatrix{ 0.4886 \cr 0.4504 \cr 0.5374 \cr 0.3256 \cr }$ 
\end{tabular}}}\parbox{2.5cm}{
\begin{equation} \end{equation}}

\centerline{\tiny \phantom{Leerzeile}}

(25) gives the normalized eigenvectors of $S^{\dagger}S$ given in the basis
of the original island modes, which is hinted at by the labels $I_i$ in
front of their components. (26) lists eigenvectors of $SS^{\dagger}$  and
their components refer to the basis of lead modes. Vectors in (26) are not
yet normalized. To do so, divide by the square root of the $\lambda_m$ above
the respective column from (25). To fully link the number examples
to the notation of the previous
sections, remark that the vectors in (25) give the columns of
$({a\!\!\!\vert}_{S^{\dagger}S})$
and those in (26) constitute $S({a\!\!\!\vert}_{S^{\dagger}S})$.
Obviously the eigenchannel system consists of four pairs of channels, denoted
as $(t_{Li},t_{Ri})$, $i=1,\ldots ,4$. With the values from (26) we shall now
work out the left and right contributions in each. For example, the
transmission amplitude of the left channel in the first pair is given by the
part of the L-modes in the normalized eigenvector multiplied by the square
root of the respective $\lambda_m$, which is the transmission amplitude
of the island eigenmode to the entire outside world, 
$t_{L1}=\sqrt{\frac{0.0627^2+0.0651^2}{0.0080}}\cdot\sqrt{0.0080}$.
The reason to give non-normalized vectors in (26) was that explicit factors
$\sqrt{\lambda_m}$ drop out here. The following table lists all left and
right transmission amplitudes as determined by (7) and (8):

\begin{center}\begin{tabular}{ccccc}
$i$ && $t_{Li}$ && $t_{Ri}$ \\ {\tiny \phantom{i}} &&&& \\
1 & \phantom{1} & $\sqrt{0.0627^2+0.0651^2}=0.0904$ & \phantom{1} &
$\sqrt{0.0026^2+0.0062^2}=0.0067$ \\
2 && $\sqrt{0.0612^2+0.0775^2}=0.0988$ &&
$\sqrt{0.0039^2+0.2050^2}=0.2050$ \\
3 && $\sqrt{0.1296^2+0.1048^2}=0.1667$ &&
$\sqrt{0.2504^2+0.0738^2}=0.2610$ \\
4 && $\sqrt{0.4886^2+0.4504^2}=0.6645$ &&
$\sqrt{0.5374^2+0.5256^2}=0.6283$ \\ 
\end{tabular}\end{center}
\begin{equation} \end{equation}

Remark that the sum of all $t_L$ squared and divided by the respective $\lambda_m$
is 2 and the same holds for the $t_R$.
\begin{equation}
\frac{0.0904^2}{0.0080}+\frac{0.0988^2}{0.0518}+\frac{0.1667^2}{0.0959}+
\frac{0.6645^2}{0.8364}=2.03
\end{equation}
\begin{equation}
\frac{0.0067^2}{0.0080}+\frac{0.2050^2}{0.0518}+\frac{0.2610^2}{0.0959}+
\frac{0.6283^2}{0.8364}=1.9992
\end{equation}
Deviations from 2 here are due to having limited accuracy to only four digits.
(28) and (29) illustrate the reason and purpose of multiplying the  
$\sqrt{\lambda_m}$ by the weights left and right sides have
in the eigenvectors. Despite four channels ending in each lead, the maximum
available supply of charges from each lead is reduced to the amount two
original modes on each side could provide. We prefer to include these weight
factors in $t_L$ and $t_R$. Alternatively, the $\sqrt{\lambda_m}$
could be called transmission amplitudes of the combined channels. However,
then a channel would not connect to equal densities of states on the left
and on the right, and weight factors for available states would have to enter
effective transfer amplitudes, anyway.

For comparison of the resulting channel transmissions let us now treat the
system as incoherent, that is as two single junctions, with the same
original scattering amplitudes.

\parbox{4.75cm}{
$$\offinterlineskip \tabskip 0pt \vbox{\halign{\strut
# \hfill & \vrule # & \vrule # &
\hfill # \hfill & \vrule # &
\hfill # \hfill & \vrule # &
\hfill # \hfill & \vrule # &
\hfill # \hfill & \vrule # \cr
$S_l$: &&& \ ${\rm i}_1$ && \ ${\rm i}_2$ && \ ${\rm i}_3$ && \ ${\rm i}_4$ & \cr
\noalign{\hrule}\noalign{\hrule}
\ $l_1$ &&& \ 0.17 && \ 0.28 && \ 0.39 && \ 0.06 & \cr
\noalign{\hrule}
\ $l_2$ &&& \ 0.06 && \ 0.22 && \ 0.40 && \ 0.11 & \cr
\noalign{\hrule}}} $$} \parbox{10cm}{\begin{equation}
$$S_lS_l^{\dagger}=\pmatrix{ 0.2630 & \ 0.2344 \cr
0.2344 & \ 0.2241 \cr }\end{equation}}

$S_lS_l^{\dagger}$
has eigenvalues 0.0083 and 0.4786, which are also the non-vanishing
eigenvalues of $S_l^{\dagger}S_l$. They correspond to transmission
amplitudes $t_{l1}=\sqrt{0.0083}=0.0911$ and $t_{l2}=\sqrt{0.4786}=0.6918$.

\parbox{4.75cm}{
$$\offinterlineskip \tabskip 0pt \vbox{\halign{\strut
# \hfill & \vrule # & \vrule # &
\hfill # \hfill & \vrule # &
\hfill # \hfill & \vrule # &
\hfill # \hfill & \vrule # &
\hfill # \hfill & \vrule # \cr
$S_r$: &&& \ ${\rm i}_1$ && \ ${\rm i}_2$ && \ ${\rm i}_3$ && \ ${\rm i}_4$ & \cr
\noalign{\hrule}\noalign{\hrule}
\ $r_1$ &&& \ 0.28 && \ 0.11 && \ 0.33 && \ 0.39 & \cr
\noalign{\hrule}
\ $r_2$ &&& \ 0.06 && \ 0.33 && \ 0.11 && \ 0.17 & \cr
\noalign{\hrule} }} $$}\parbox{10cm}{\begin{equation}
S_rS_r^{\dagger}=\pmatrix{ 0.3515 & \ 0.1557 \cr
0.1557 & \ 0.1535 \cr } \end{equation}}

Eigenvalues of $S_rS_r^{\dagger}$ as well as of $S_r^{\dagger}S_r$ are
0.0700 and 0.4370 and these translate
into transmission amplitudes $t_{r1}=\sqrt{0.0700}=0.2646$ and 
$t_{r2}=\sqrt{0.4370}=0.6610$. Like for the single junction the number
of channels in each junction is equal to the smaller number of modes on
one side of a junction. There are two channels per junction here. These
may, however, have four different linear combinations of original
island modes as new modes at their ends on the island. The transmission
amplitudes are comparable in size to some of those found in the coherent case,
mostly the greater ones for a junction, but there is no easy way of
predicting the channel ensembles for the coherent double junction from
those for both junctions taken as singles.

In order to verify that modes only coupled across one junction and disconnected
from a network extending over both result in channels through only one
junction, we have done an analogous calculation to the above one
starting with the following scattering matrix:

\parbox{8cm}{
$$\offinterlineskip \tabskip 0pt \vbox{\halign{\strut
# \hfill & \vrule # & \vrule # &
\hfill # \hfill & \vrule # &
\hfill # \hfill & \vrule # &
\hfill # \hfill & \vrule # &
\hfill # \hfill & \vrule # \cr
$S$: &&& \ $I_1$ && \ $I_2$ && \ $I_3$ && \ $I_4$ & \cr
\noalign{\hrule}\noalign{\hrule}
\ $L_1$ &&& \ 0.17 && \ 0.28 && \ 0.39 && \ 0 & \cr
\noalign{\hrule}
\ $L_2$ &&& \ 0.06 && \ 0.22 && \ 0.40 && \ 0 & \cr
\noalign{\hrule}
\ $R_1$ &&& \ 0.28 && \ 0.11 && \ 0.33 && \ 0 & \cr
\noalign{\hrule}
\ $R_2$ &&& \ 0 && \ 0 && \ 0 && \ 0.17 & \cr
\noalign{\hrule} }} $$}\parbox{3.5cm}{
\begin{tabular}{ccccc}
$i$ & \phantom{1}  & $t_{Li}$ &\phantom{1}& $t_{Ri}$  \\
1 && 0.0606 && 0.0161 \\
2 && 0 && 0.17 \\
3 && 0.1087 && 0.1497 \\
4 && 0.6752 && 0.4204 
\end{tabular}}\parbox{1cm}{\begin{equation} \end{equation}}

This results in three eigenvectors the fourth component of which vanishes
(eigenvalues 0.0039, 0.0342 and 0.6326) and one eigenvector of which the
first three are zero (eigenvalue 0.0289). The resulting
transmission amplitudes of channel pairs are listed in (32).
The second line with $t_{L2}=0$ obviously represents a single channel
only bridging the right junction. The input weight of left modes is 2,
\begin{equation}
\frac{0.0606^2}{0.0039}+0+\frac{0.1087^2}{0.0342}+\frac{0.6752^2}{0.6326}=2.01\ .
\end{equation}
The input weight of all right modes is also 2,
\begin{equation}
\frac{0.0161^2}{0.0039}+\frac{0.17^2}{0.0289}+\frac{0.1497^2}{0.0342}+
\frac{0.4204^2}{0.6326}=2.001\ ,
\end{equation}
where 1 falls to the single channel with $t_{R2}=0.17$ and the others amount to
1 altogether.

The following example is made up for a qualitative estimation on how a junction
that exhibits a channel with medium transmission taken as stand-alone
device and a tunnel junction that has many low-transmission channels would behave
when combined with coherent coupling across the island between them.
To make the system as coherent as possible except for paths through the tunnel
junction, we assume that many modes from the right lead couple weakly
to the same mode on the island. With one mode on the left, one on the island
and modes $R_1$
to $R_{10}$ on the right let the original scattering amplitudes be given by
\begin{equation}
\offinterlineskip \tabskip 0pt \vbox{\halign{\strut
# \hfill & \vrule # & \vrule # &
\hfill # \hfill & \vrule # &
\hfill # \hfill & \vrule # &
\hfill # \hfill & \vrule # &
\hfill # \hfill & \vrule # &
\hfill # \hfill & \vrule # \cr
$S^{\dagger}$: &&& \ $L$ && \ $R_1$ && \ $R_2$ && \ $\ldots$ && \ $R_{10}$ & \cr
\noalign{\hrule}\noalign{\hrule}
$I$ &&& \ 0.5 && \ 0.01 && \ 0.01 && \ \ldots && \ 0.01 & \cr
\noalign{\hrule}}} \end{equation}
Then $S^{\dagger}S=0.5^2+10\cdot 0.01^2=0.251$
and there is only one non-zero $\sqrt{\lambda}=\sqrt{0.251}=0.501$. The
associated eigenvector of $S^{\dagger}S$ is {\it (1)}, that is the island
mode is the eigenmode. It translates into the
combination $0.5L+0.01R_1+0.01R_2+\ldots +0.01R_{10}$ in the lead basis.
The system exhibits one channel pair with $t_L=\sqrt{0.5^2}=0.5$ and
$t_R=\sqrt{10\cdot 0.01^2}=0.032$. From the right only the eigenmode
$(0.01R_1+0.01R_2+\ldots +0.01R_{10})/\sqrt{10\cdot 0.01^2}$ 
is transmitted, all other linear combinations of right modes are not.
The tunnel-junction appears as one effective channel the transmission
amplitude of which is enhanced over a single original path by the square root
of the number of paths having been assumed with equal throughput.
Another example with unequal numbers of original modes left and right will be
contained in the next section.
There is a tendency of left and right channel transmissions being paired
ordered by size. This is no principal necessity, however. For example,
the channel left with the largest (smallest) $t_L$ need not be coherently
connected to the channel right with the largest (smallest) $t_R$.

The last example in this section is intended to illustrate how subtle the
differences between a valid scattering matrix and a matrix violating
probability conservation can be, especially in comparing double to single
junctions. Condition (3) is a sufficient criterion, however, set up
overcautiously, and thus not a necessary one. We have already seen that if
we were to take the matrix from (9) for a single junction between L and I,
for example,
\begin{equation}
\offinterlineskip \tabskip 0pt \vbox{\halign{\strut
# \hfill & \vrule # & \vrule # &
\hfill # \hfill & \vrule # &
\hfill # \hfill & \vrule # &
\hfill # \hfill & \vrule # &
\hfill # \hfill & \vrule # \cr
$S_l$: &&& \ ${\rm i}_1$ && \ ${\rm i}_2$ && \ ${\rm i}_3$ && \ ${\rm i}_4$ & \cr
\noalign{\hrule}\noalign{\hrule}
$l_1$: &&& \ 0.3 && \ 0.5 && \ 0.5 && \ 0.6 & \cr
\noalign{\hrule}
$l_2$: &&& \ 0.1 && \ 0.4 && \ 0.8 && \ 0.2 & \cr 
\noalign{\hrule}\noalign{\hrule}}}
\end{equation}
although the norm of each line and each column vector is less than 1,
this would result in an eigenvalue and thus a transmission amplitude for
one of the two channels $t_2$=$\sqrt{\lambda_2}$=1.285 greater than one.
Supposing that the first line gives scattering of four island modes to a
single mode on the left, again for only a single junction between L and I,

\parbox{7.5cm}{$$
\offinterlineskip \tabskip 0pt \vbox{\halign{\strut
# \hfill & \vrule # & \vrule # &
\hfill # \hfill & \vrule # &
\hfill # \hfill & \vrule # &
\hfill # \hfill & \vrule # &
\hfill # \hfill & \vrule # \cr
$S_l$: &&& \ ${\rm i}_1$ && \ ${\rm i}_2$ && \ ${\rm i}_3$ && \ ${\rm i}_4$ & \cr
\noalign{\hrule}\noalign{\hrule}
$l$: &&& \ 0.3 && \ 0.5 && \ 0.5 && \ 0.6 & \cr 
\noalign{\hrule}\noalign{\hrule}}}$$}\parbox{7cm}{\begin{equation}
S_lS_l^{\dagger}=\pmatrix{ 0.95 } \end{equation}}

then it is easily seen from $S_lS_l^{\dagger}$ that the only non-vanishing
eigenvalue is $\lambda$=0.95 corresponding to $t$=$\sqrt{0.95}$=0.975.
Analogously supposing that the second line respresents scattering through
a single junction, named as between island and right,
\newline {\tiny \phantom{nix}}

\parbox{7.5cm}{$$
\offinterlineskip \tabskip 0pt \vbox{\halign{\strut
# \hfill & \vrule # & \vrule # &
\hfill # \hfill & \vrule # &
\hfill # \hfill & \vrule # &
\hfill # \hfill & \vrule # &
\hfill # \hfill & \vrule # \cr
$S_r$: &&& \ ${\rm i}_1$ && \ ${\rm i}_2$ && \ ${\rm i}_3$ && \ ${\rm i}_4$ & \cr
\noalign{\hrule}\noalign{\hrule}
$r$: &&& \ 0.1 && \ 0.4 && \ 0.8 && \ 0.2 & \cr 
\noalign{\hrule}\noalign{\hrule}}}$$}\parbox{7cm}{\begin{equation}
S_rS_r^{\dagger}=\pmatrix{ 0.85 } \end{equation}}

the eigenvalue and the transmission amplitude for the only channel here
would be $\lambda=0.85$ and $t$=0.922. If now we were to take the matrix (9)
as set up for the double junction with one mode left and one right
\newline {\tiny \phantom{nix}}

\parbox{6cm}{$$
\offinterlineskip \tabskip 0pt \vbox{\halign{\strut
# \hfill & \vrule # & \vrule # &
\hfill # \hfill & \vrule # &
\hfill # \hfill & \vrule # &
\hfill # \hfill & \vrule # &
\hfill # \hfill & \vrule # \cr
$S$: &&& \ $I_1$ && \ $I_2$ && \ $I_3$ && \ $I_4$ & \cr
\noalign{\hrule}\noalign{\hrule}
$L$: &&& \ 0.3 && \ 0.5 && \ 0.5 && \ 0.6 & \cr
\noalign{\hrule}
$R$: &&& \ 0.1 && \ 0.4 && \ 0.8 && \ 0.2 & \cr 
\noalign{\hrule}\noalign{\hrule}}}$$}
\parbox{5.5cm}{\begin{tabular}{ccc}
$\lambda$:  & 0.148 & 1.652 \\ {\tiny \phantom{i}} && \\
$\matrix{ L: \cr R: \cr}$ & $\pmatrix{0.6832 \cr -0.7302}$ &
$\pmatrix{-0.7302 \cr -0.6832}$ \\ \end{tabular}
}\parbox{1cm}{\begin{equation}\end{equation}}

\vspace{0.1cm}

then from the corresponding eigenvectors of $SS^{\dagger}$
associated to $\lambda_1$=0.148 and $\lambda_2$=1.652 we get as transmission
amplitudes

\begin{center}\begin{tabular}{ccccc}
$i$ & \phantom{i} & $t_{Li}$ & \phantom{i} & $t_{Ri}$ \\
{\tiny \phantom{i}} &&&& \\
1 && $0.6832\cdot\sqrt{0.148}=0.2628$ && $0.7302\cdot\sqrt{0.148}=0.2809$ \\
2 && $0.7302\cdot\sqrt{1.652}=0.9385$ && $0.6832\cdot\sqrt{1.652}=0.8781$ \\
\end{tabular}
\end{center}

and total mode weights left and right
\begin{eqnarray}
0.2628^2/0.148+0.9385^2/1.652&=&1.0 \\
0.2809^2/0.148+0.8781^2/1.652&=&1.0
\end{eqnarray}
Even though a $\lambda$ exceeds one, no $t$ does. (36) is no allowed
scattering matrix for a single junction, but (39) from the result for all
$t$ looks possible for a double junction. 
(We should nevertheless have rejected this $S$, because (13) tells us that
one backscattering off the island of the two lead modes amounts to a weight
of more than two, which indicates an overlap situation as in Fig.4b.)
These considerations show once more that 
the double junction has to be treated as a principally different situation
and its quantitative properties cannot be deduced in a simple manner from
those of the single junctions.

For the algorithms presented here, that is for diagonalizing a matrix
$S^{\dagger}S$ and determining its eigenvalues and even for splitting up
such an eigenvalue into amplitudes $t_L$ and $t_R$ for a pair of channels
in the double junction, it is of no importance whether the eigenvalues
$\lambda$ are less or equal to one or not. You could be tempted to interpret
a $t>1$ as summing configurationally degenerate modes in a reservoir that
quantitatively have identical overlaps with all modes on the other side of
a junction and together are well supplied with charge carriers with a weight
two, three, etc. instead of one. Such modes should however be entered into
$S$ as several, albeit identical, lines or columns to ensure for the
orthogonality of all modes and the premises of each being fed with weight one.
Transmission amplitudes $t$ greater than one are unphysical.
Green's functions \cite{cuevgreen,my06,tdyson}
that represent transmissions renormalized for multiple
reflections require single-hopping amplitudes less or equal to one per
transport channel. \footnote{The easiest case is $T=t/(1+t^2)$ for a
channel in a single junction in the normal conducting case, which stems
from a geometric series \cite{arxiv}.}

Of course, being given scattering matrices for two individual junctions -
taking reflection parts with - there is a straight forward rule how to combine
two such matrices into one describing the two junctions put in series (with
what we call coherent coupling) \cite{datta}. So there has to be a reason
why we reopen the problem from the point of view of eigenchannels and newly
invent a way for describing the double junction. The fact that for a single
junction from experimental results eigenchannel ensembles can be deduced
\cite{elke}, whereas full scattering matrices cannot, does not lead to an
irrefutable justification. Merely being given current-voltage characteristics,
for the double junction there may even be some ambiguity in deciding whether
transport is coherent or incoherent \cite{my07}. Like scattering matrices our
concept with channel pairs primarily provides a theoretical ansatz for
modelling. Convoluting scattering matrices of two junctions in series into
one effective matrix once and for all \cite{datta} would be appropriate
if the island between the junctions did not change its state through charge
transport onto and off it (such a resulting matrix would mimic an effective
single junction). However, in the situation we are interested in, prone to
Coulomb blockade, the island potential changes with each charge transfer and
thus the energy range of available electron or hole states does, too. Thus
an all-inclusive scattering matrix or a renormalized transfer Green's function
has to depend on all possible island charge states. Convoluting here scattering
matrices or Green's functions of two junctions into effective global matrices
or functions, with the further necessity to transform to Fourier space in
order to account for energy-dependent densities of states, is practically
impossible for cross-linked transfer paths as in Fig.3a, whereas this is
feasible for paired channels as in Fig.3b \cite{tdyson}. Besides that, the
original transfer Green's functions approach to transport properties of
especially a superconducting quantum point contact \cite{cuevgreen} was made
without separating modes into ingoing and outcoming parts and with restricting
scattering matrices to transmission parts (called hopping elements). An
extension of that formalism to double junctions should keep these premises.

\section{Partial coherence}

With today's microstructering techniques, particularly with metals, it is
possible to fabricate islands that are still large enough to exhibit a
bulk-like continuous density of states, however already small enough to be
sensitive to single-electron charging \footnote{Island dimensions, for
example for aluminum, are typically from below one up to two micrometers
across and some ten nanometers in height.} \cite{SET}.
One may further presume a mixture of
coherent and incoherent interaction between processes in both junctions. It
shall be shown that such partially coherent transport can be described by
finding a system of channels following exactly the same method we just
presented for the coherent case. In fact, the system will look like a
coherent one with channels of just little lower transmission than one set up
with coherent coupling across the island only.

Also the fully coherent case allows sequential (incoherent)
transport. A coherent multiscattering process can end on the island letting
an electron or hole having come in into any mode there relax into the
reservoir of charge carriers equally populating all modes. Another process,
that decharges the island and brings it back to its original potential, may
then begin out of any mode on the island.

To change from full coherence to partial coherence, we want
to reduce coherence in transport across the island for a charge carrier
coming into a mode there from a lead, but keep the coupling of that island
mode to each lead the same, that is not decrease the transmissions of the
junctions, an idea reminiscent of \cite{buettpartial}. It can here
be done by using the following picture. Replace the
original island mode which was coupled to modes at L and R by a threefold
of new modes: One that is coupled to L- and R-modes as before, but with
all scattering amplitudes reduced by a factor of $\sqrt{p}$, one that is
coupled only to the L-modes and one that is coupled only to the R-modes.
As scattering amplitudes of the latter two new I-modes take the original
amplitudes reduced by a factor of $\sqrt{1-p}$. $p$ is number between 0 and 1
and can be interpreted as the probability that a charge carrier entering into
the chosen island mode gets coherently transported across to the other junction.
The sum of absolute squares of scattering out of the island mode stays the
same as before. For L- or R-modes only their properties localized at a
junction count. As compared to the single junction, in contrast to the
island in the leads there is no newly introduced specific distance, over
which it is important to know to what degree coherence is maintained or lost.
For lead modes a splitting as done for island modes does not make sense.
Of course, lead and island modes only coupled across one junction and in
no way connected to a network coherently linking them to the other lead
may always also be present from the beginning.

\begin{figure*}
\parbox{7cm}{
\includegraphics[width=7cm]{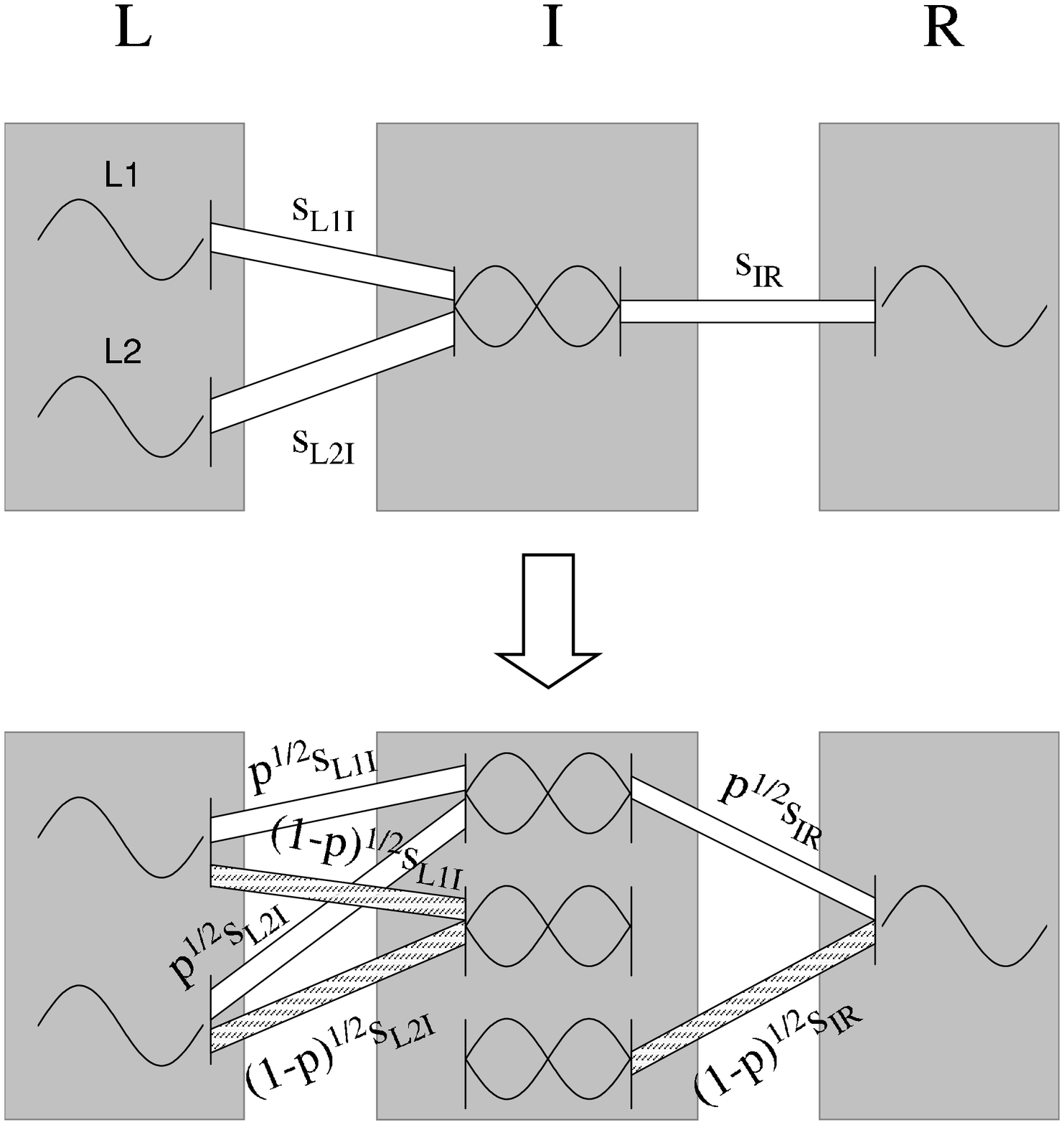}}
\parbox{1cm}{\phantom{nix}}
\parbox{7cm}{
{\small {\bf Figure 5.} An original island mode with couplings as set up for the
fully coherent model is replaced by three new modes coupling to both or only
one lead, respectively.}}
\end{figure*}

\setcounter{figure}{5}

For comparison with partial coherence we first elaborate the following
fully coherent case with two modes on the island, two left and one right.
\parbox{7cm}{$$
\offinterlineskip \tabskip 0pt \vbox{\halign{\strut
# \hfill & \vrule # & \vrule # &
\hfill # \hfill & \vrule # &
\hfill # \hfill & \vrule # \cr
$S$: &&& \ $I_1$ && \ $I_2$ &  \cr
\noalign{\hrule}\noalign{\hrule}
\ $L_1$ &&& \ 0.17 && \ 0.30 & \cr \noalign{\hrule}
\ $L_2$ &&& \ 0.06 && \ 0.40 & \cr \noalign{\hrule}
\ $R$ &&& \ 0.28 && \ 0.35 & \cr \noalign{\hrule}}}$$}
\parbox{7cm}{\begin{equation}
S^{\dagger}S=\pmatrix{ 0.1109 & \ 0.1730 \cr 0.1730 & \ 0.3725 \cr} \quad
\end{equation}}

$S^{\dagger}S$ has eigenvalues with associated eigenvectors:

\parbox{10cm}{
\begin{center}\begin{tabular}{ccc}
0.0248 & 0.4586 \\ {\tiny \phantom{nix}} & \\
$\pmatrix{ 0.8953 \cr -0.4455 \cr }$ &
$\pmatrix{ 0.4455 \cr 0.8953 \cr }$ &
$\matrix{ \leftarrow I_1 \cr \leftarrow I_2\cr }$
\end{tabular}\end{center}}
\parbox{5cm}{\begin{equation} \end{equation}}

Multiplying by $S$ translates them into eigenvectors of $SS^{\dagger}$:
\begin{equation}
\pmatrix{0.17 & 0.30 \cr 0.06 & 0.40 \cr 0.28 & 0.35 }
\pmatrix{0.8953 \cr -0.4455 \cr }
=\pmatrix{ 0.0186 \cr -0.1245 \cr 0.0948 \cr } \; \;
\matrix{ \leftarrow L_1 \cr \leftarrow L_2 \cr \leftarrow R \cr }
\end{equation}
\begin{equation}
\pmatrix{ 0.17 & 0.30 \cr 0.06 & 0.40 \cr 0.28 & 0.35 \cr }
\pmatrix{ 0.4455\cr 0.8953\cr }=
\pmatrix{ 0.3443\cr 0.3849\cr 0.4381\cr }
\matrix{ \leftarrow L_1 \cr \leftarrow L_2 \cr \leftarrow R \cr }
\end{equation}
From left and right parts in these we read off the transmission amplitudes
of two pairs of channels:

\begin{center}\begin{tabular}{ccccc}
$i$ & \phantom{1} & $t_{Li}$ &\phantom{1}& $t_{Ri}$ \\
{\tiny \phantom{nix}} &&&& \\
1 && $\sqrt{0.0186^2+0.1245^2}$=0.1259 && 0.0948 \\
2 && $\sqrt{0.3443^2+0.3845^2}$=0.5164 && 0.4381
\end{tabular}\end{center}

In this example remark that the input on the left having a chance to be
transmitted amounts to less than 2 modes and that on the right to less than 1.
\begin{equation} 0.1259^2/0.0248+0.5164^2/0.4586=1.22 \end{equation}
\begin{equation} 0.0948^2/0.0248+0.4381^2/0.4586=0.78 \end{equation}
Together they have the weight of 2 modes. The island only offers 2 modes.
Thus only a two-dimensional subspace of the three-dimensional vector space
spanned by all lead modes can contribute to transmission through the system.

Let us now suppose that the $I_2$-mode is split into parts coherently
transferred between the junctions and parts losing coherence on the island.
The three modes replacing
$I_2$ are $I_{2c}$ for the coherent part and $I_{2L}$ and $I_{2R}$ for the
parts coupling to the left and right lead only, respectively. We choose
$p=0.5$. In this special case, $1-p=0.5$, too. Compared to column $I_2$ from
$S$ in (42) non-zero numbers in columns $I_{2c}$, $I_{2L}$ and $I_{2R}$
here are multiplied by $\sqrt{0.5}$, for example, $0.30\cdot \sqrt{0.5}$=0.212.

\vspace{0.2cm}

\parbox{7.5cm}{
$\offinterlineskip \tabskip 0pt \vbox{\halign{\strut
# \hfill & \vrule # & \vrule # &
\hfill # \hfill & \vrule # &
\hfill # \hfill & \vrule # &
\hfill # \hfill & \vrule # &
\hfill # \hfill & \vrule # \cr
$S$: &&& \ $I_1$ && \ $I_{2c}$ && \ $I_{2L}$ && \ $I_{2R}$ & \cr
\noalign{\hrule}\noalign{\hrule}
\ $L_1$ &&& \ 0.17 && \ 0.212 && \ 0.212 && 0 & \cr
\noalign{\hrule}
\ $L_2$ &&& \ 0.06 && \ 0.283 && \ 0.283 && 0 & \cr
\noalign{\hrule}
\ $R$ &&&  \ 0.28 && \ 0.247 && 0 && \ 0.247 & \cr
\noalign{\hrule}}}$} \parbox{8cm}{
$S^{\dagger}S=\pmatrix{
0.1190 & \ 0.1222 & \ 0.0530 & \ 0.0692 \cr
0.1222 & \ 0.1860 & \ 0.1250 & \ 0.0610 \cr
0.0530 & \ 0.1250 & \ 0.1250 & \ 0 \cr
0.0692 & \ 0.0610 & \ 0 & \ 0.0610 \cr }$
\begin{equation} \end{equation}}

Because there is a lead mode less than island modes, one of the eigenvalues,
noted below together with the corresponding eigenvectors, is zero.

\parbox{12cm}{
\begin{center}\begin{tabular}{ccccc}
& 0 & 0.0059 & 0.1040 & 0.3730 \cr {\tiny \phantom{nix}} &&&& \\
$\matrix{ I_1: \cr I_{2c}: \cr I_{2L}: \cr I_{2R}: \cr }$ &
$\pmatrix{ 0 \cr -0.5774 \cr 0.5774 \cr 0.5774 \cr }$ &
$\pmatrix{ -0.7583 \cr 0.4085 \cr -0.0912 \cr 0.4997 \cr }$ &
$\pmatrix{ -0.4353 \cr 0.0723 \cr 0.6697 \cr -0.5974 \cr }$ &
$\pmatrix{ 0.4852 \cr 0.7033 \cr 0.4581 \cr 0.2451 \cr }$
\end{tabular}\end{center}}
\parbox{2.5cm}{\begin{equation} \end{equation}}

If now we multiply these by $S$ we obtain the (non-normalized)
eigenvectors of $SS^{\dagger}$,
noted again below the corresponding eigenvalues here:

\parbox{12cm}{
\begin{center}\begin{tabular}{ccccc}
& 0 & 0.0059 & 0.1040 & 0.3730 \\ {\tiny \phantom{nix}} &&&& \\
$\matrix{ L_1: \cr L_2: \cr R: \cr}$ &
$\pmatrix{ 0 \cr 0 \cr 0 \cr }$ &
$\pmatrix{ -0.0616 \cr 0.0443 \cr 0.0120 \cr }$ &
$\pmatrix{ 0.0833 \cr 0.1839 \cr -0.2516 \cr }$ &
$\pmatrix{ 0.3287 \cr 0.3578 \cr 0.3701 \cr }$ 
\end{tabular}\end{center}}\parbox{2.5cm}{
\begin{equation} \end{equation}}

The zero-vector here indicates that there is a combination of island modes
$I_1$, $I_{2c}$, $I_{2L}$ and $I_{2R}$, namely that given by the first vector
in the above list of four-vectors, which is never transmitted.
This state cannot exchange charge
carriers with the lead modes. From the other three eigenvectors we read off
the transmission amplitudes of three pairs of channels:

\begin{center} \begin{tabular}{ccccc}
$i$ & \phantom{1} & $t_{Li}$ & \phantom{2} & $t_{Ri}$ \\
{\tiny \phantom{i}} & & & \\ 
$1$ && $\sqrt{0.0616^2+0.0443^2}$=0.0759 && 0.0120 \\
$2$ && $\sqrt{0.0833^2+0.1839^2}$=0.2019 && 0.2516 \\
$3$ && $\sqrt{0.3287^2+0.3578^2}$=0.4859 && 0.3701
\end{tabular}\end{center}

Here again, the maximum possible input left amounts to two modes, and that
right to one:
\begin{eqnarray}
0.0759^2/0.0059+0.2019^2/0.1040+0.4859^2/0.3730 &=& 2\\
0.0120^2/0.0059+0.2516^2/0.1040+0.3701^2/0.3730 &=& 1 
\end{eqnarray}
Channels are not at all the same as when letting the $I_2$-mode being
transported fully coherently across the island. The important finding is,
however, that, even if by construction some of the original island modes
only have a direct overlap with one lead, the eigenchannel
ensemble does not contain channels merely bridging a single junction
(if no paths are totally separated from the rest of the original network).
The connection of the coherently and incoherently transported
island modes via the lead modes causes the system to appear as effectively
consisting of fully coherently linked channel pairs. As already pointed out
earlier, the system of channel pairs does not suppress
sequential transport. The latter may well be found to provide the dominating
contributions in current-voltage characteristics.
The channels with the largest throughput left and right may form a pair,
in principle, however, as well 
belong to different pairs \footnote{With coherent coupling, transmission
probabilities $\theta$ cannot be inferred from single $t$ alone.
A renormalization is analytically possible in the normal conducting case,
and a channel on the left, for example, would get $\theta_{Li}=4t_{Li}^2/(
1+t_{Li}^2+t_{Ri}^2)^2$ \cite{negrate}.}.

\section{Conclusions}

It has been discussed how to determine the ensemble of transport channels
for a series of two point contacts enclosing an island
between them, which allows coherent
transport across it. We find that transport channels are only coherently
coupled together in pairs involving one channel per junction. This makes
a Green's functions algorithm developed for single-channel junctions
coherently coupled via an island \cite{my07} applicable to the general case
of multichannel junctions. The channel ensembles describing coherently
coupled contacts will differ from those describing the same contacts each
taken as a stand-alone device. Partially coherent transport across the
island can also be treated with the presented method of determining transport
channels and will effectively look like coherent transport through more but
less open channels. Circuits with more than two contacts in series or
with short-cutting channels between leads,
however, would be more complicated to handle.


\vspace{1cm}

\begin{thebibliography}{14}

\bibitem{Grabert} Grabert H and Devoret M {\it Single Charge Tunneling}
New York 1992

\bibitem{Buett} B\"uttiker M and Polianski M L 2005 
{\it J.Phys.A: Math.Gen.} {\bf 38} 10559

\bibitem{imry} Imry Y {\it Introduction to Mesoscopic Physics}
Oxford 2002

\bibitem{beenakker} Beenakker C W J and van Houten H 1991, {\it Solid State
Physics} {\bf 44} 1

\bibitem{datta} Datta S {\it Electronic Transport in Mesoscopic Systems}
Cambridge 2001

\bibitem{quantcond}
van Wees B J, van Houten H, Beenakker C W J, Williamson J G and Foxon C T
1988 {\it Phys.Rev.Lett.} {\bf 60} 848;
Crook R, Prance J, Thomas K J, Chorley S J, Farrer I, Ritchie D A, Pepper M
and Smith C G 2006 {\it Science} {\bf 312} 135. 
There are deviations from integer conductances in long junctions:
Gloos K, Utko P, Aagesen M, S\o rensen C B, Bindslev Hansen J and Lindelof P E
2006 {\it Phys.Rev.B} {\bf 73} 125326.
And anomalies due to spin-splitting: Bychkov A M and Stace T M 2007
{\it Nanotechnolgy} {\bf 18} 185403.

\bibitem{elke} Scheer E {\it et al.} 1998 {\it Nature} {\bf 394} 154;
Scheer {\it et al.} 1997 {\it Phys.Rev.Lett.} {\bf 78} 3535.

\bibitem{cuevmicro} Cuevas J C, Levy Yeyati A and Martin-Rodero A 1998
{\it Phys.Rev.Lett.} {\bf 80} 1066.

\bibitem{cuevgreen} Cuevas J C, Martin-Rodero A and Levy Yeyati A 1996
{\it Phys.Rev.B} {\bf 54} 7366.

\bibitem{my06} Schr\"oter U and Scheer E 2006 {\it Phys.Rev.B} {\bf 74} 245301. 

\bibitem{my07} Schr\"oter U and Scheer E 2007 {\it Phys.Rev.B} {\bf 76} 205104.

\bibitem{tdyson} Schr\"oter U and Scheer E 2008 {\it J.Phys.A: Math.Theor.}
{\bf 41} 265202.

\bibitem{arxiv} Schr\"oter U {\it cond-mat} arXiv:0802.3001

\bibitem{SET} Refer to metallic single-electron transistors, however,
contacts there are tunnel junctions. For example:
Hadley P, Delvigne E , Visscher E H, L\"ahteenm\"aki S and Mooij J E 1998
{\it Phys.Rev.B} {\bf 58} 15317;
Hergenrother J M, Tuominen M T and Tinkham M 1994 {\it Phys.Rev.Lett.}
{\bf 72} 1742;
Billangeon P M, Pierre F, Bouchiat H and Deblock R 2007 {\it Phys.Rev.Lett.}
{\bf 98} 216802.

\bibitem{buettpartial} B\"uttiker M 1988 {\it IBM J.Res.Develop.} {\bf 32} 63

\bibitem{negrate} Schr\"oter U and Scheer E 2008 {\it J.Phys.A: Math.Theor.}
{\bf 41} 375203.

\end{thebibliography}
\end{document}